\documentclass[pre,showpacs,amssymb,preprint,nofootinbib]{revtex4}

\usepackage{epsfig}
\usepackage{wasysym}
\usepackage{ae}
\usepackage[T1]{fontenc}
\usepackage{bbold,psfrag}
\usepackage{color}

\newcommand{\di}{\displaystyle}
\newcommand{\mc}{\mathcal}
\newcommand{\be}{\begin{equation}}
\newcommand{\ee}{\end{equation}}
\newcommand{\bd}{\begin{displaymath}}
\newcommand{\ed}{\begin{displaymath}}
\newcommand{\bea}{\begin{eqnarray}}
\newcommand{\eea}{\end{eqnarray}}
\newcommand{\wh}{\widehat}

\begin{document}
\title{Anisotropies in thermal Casimir interactions: ellipsoidal colloids trapped at a fluid interface}
\date{\today}
\author{Ehsan Noruzifar}
\author{Martin Oettel}
\affiliation{Institut f\"ur Physik, WA 331, Johannes-Gutenberg-Universit\"at
Mainz,
55099 Mainz, Germany}
\begin{abstract}
We study  the effective interaction between  two ellipsoidal particles
 at the interface of two fluid phases
which are  mediated by thermal fluctuations of the interface. 
In this system the restriction of the long--ranged
 interface fluctuations by particles 
gives rise to fluctuation--induced forces which are
 equivalent to interactions of Casimir type and which are anisotropic
in the interface plane. 
Since the position and the orientation of the colloids with respect to the
interface normal may also fluctuate, this system is an example for the
Casimir effect with fluctuating boundary conditions.  
In the approach taken here, the Casimir interaction is rewritten
 as the interaction between fluctuating multipole
moments of an auxiliary charge density--like field defined on the area enclosed by the contact lines.
These fluctuations are coupled to fluctuations of multipole moments of the contact line position
(due to the possible position and orientational fluctuations  of the colloids).
We obtain explicit expressions for the behavior of the Casimir interaction at large distances
for arbitrary ellipsoid aspect ratios.
If colloid fluctuations are suppressed, the Casimir interaction at large distances is 
isotropic, attractive and long ranged (double--logarithmic in the distance). 
If, however, colloid fluctuations are included, the Casimir interaction at large distances
changes to a power law in the inverse distance and becomes anisotropic.
The leading power is 4 if only vertical fluctuations of the colloid center are allowed,
and it becomes 8 if also orientational fluctuations are included.
\end{abstract}

\maketitle
%

\section{Introduction}\label{sec:intro}
When a fluctuating medium with long--ranged, power--law  correlations is confined between
a set of boundaries, forces with likewise long--ranged character are induced between the boundaries.
{ There are different possible  sources of such fluctuations in a 
medium: in a quantum--mechanical system it is the zero point energy of the vacuum 
(or ground state), 
and in a classical system it is the finite temperature which causes order parameter
fluctuations \cite{kargol_modrev}.
}

This kind of force was discovered theoretically by Casimir in 1948
 for the case of two parallel,  conductive and uncharged plates immersed in vacuum 
which he attributed to
zero--point fluctuations of the electromagnetic field \cite{casimir}. 
For a review on recent progress and the status of experimental verification of this 
quantum mechanical Casimir effect, see Ref.~\cite{Mil04}.
A classical equivalent of the Casimir force observable between objects immersed in 
a fluid in the vicinity of its critical point was predicted 30 years later \cite{degennes}. 
{The fluctuations of the order parameter field near the critical point are long--ranged,
and thus they
give rise to a Casimir--like, fluctuation-induced force.}
This effect has recently been observed in an experiment probing the force on colloidal particles 
which have been immersed
in a near--critical binary mixture in the vicinity
of a wall \cite{Gam08}.
Another classical variant of the Casimir interaction is found
between particles (colloids) that are trapped at membranes \cite{goulian,rods} or at the interface of 
two fluid phases \cite{cw}.
 In this two--dimensional latter instance, thermally excited height fluctuations 
of the interface which have long--ranged nature are disturbed by 
the presence of colloids. The energy spectrum of the fluid interface
on a coarse--grained level is very well described 
by an effective capillary wave Hamiltonian which governs both 
the equilibrium interface configuration and the thermal fluctuations
 around this equilibrium. Since capillary waves are 
the Goldstone modes of the broken translational symmetry pertaining 
to a free interface, their correlations decay logarithmically in 
the absence of gravity and the corresponding fluctuation--induced 
forces are a manifestation of the Casimir effect for a Gaussian theory in two 
dimensions.
{Compared to manifestations of the Casimir effect in a three-dimensional bulk medium,
a new phenomenon arises here: The boundary of the fluctuating interface, which is
represented by the contact line on the colloid surface, is itself mobile due to
position fluctuations of the colloids and finite surface tensions of the colloid--liquid
interfaces and thus the Casimir force receives another contribution due to these
fluctuating boundaries.   
This effect has been noticed first for colloidal rods trapped at membranes and films
\cite{rods}.
  For a system composed of two spherical colloids
 trapped at a fluid interface,
various realizations of these fluctuating boundary conditions are possible and
it has been shown \cite{martin1, martin2, martin3} that the fluctuation-induced force sensitively depends on the
type of boundary conditions, with the asymptotics of the force ranging from $1/(d\ln d)$ to 
$d^{-9}$.}
 
{In recent work, the general form of the fluctuation-induced interactions between a finite
 number of compact objects of arbitrary shapes and separations has been calculated
for a fluctuating medium of scalar Gaussian type \cite{jaffe1} and an electromagnetic
medium \cite{jaffe2} with fixed boundary conditions on the surface of the objects. 
This has been achieved by viewing the Casimir interaction as resulting from fluctuations
of source distributions (of the fluctuating field) on the surfaces of the objects
which are decomposed in terms of multipoles.} 
Then by a functional integral over the effective action of multipoles,
 the resulting interaction has been found. 
{In such a way, the effect of anisotropic object shape on the Casimir force appears 
to be tractable.}
However, the objects' shape and the boundary conditions enter the 
effective action by its scattering matrix which is a nontrivial object already for simple shapes
\cite{jaffe1}. Studies with explicit calculations for objects other than spheres, cylinders and walls 
have partially  been focused 
on the effect of wall corrugations \cite{Bue04,Gie08}, but also sharp edges \cite{Gie06} and 
rectangular bodies \cite{Mil08}
have been investigated. 
{Also, in a recent work, the quantum Casimir interaction between two 
ellipsoidal particles as well as an ellipsoidal particle and a plane has been studied \cite{emig09}.}
In the present work, we investigate the fluctuation--induced 
interactions between two ellipsoidal colloids that are trapped
 at the interface of two fluid phases {with special emphasis on the effect of anisotropy. 
Ellipsoids (spheroids) of varying aspect ration allow a smooth interpolation between
spheres \cite{martin1} and rods \cite{rods}. } The ellipsoidal colloids 
are assumed to be  of Janus type, therefore the interface contact line 
is always pinned to the colloids surface, nevertheless the vertical 
position of colloids and their orientation may fluctuate, {giving rise to the already mentioned
feature of fluctuating boundaries. For the calculation of the Casimir force, we employ techniques
which have been introduced in previous work \cite{martin1, martin2, rods} and which partially can also be
interpreted in terms of the scattering matrix ansatz of Refs.~\cite{jaffe1,jaffe2} such that our results
are a concrete example of the general theory for the effects of anisotropic object shape.
As stressed before, however, at an interface we always have the influence of the fluctuating boundary 
conditions. 
In order to study the effect of the mobile boundaries in our work, 
we have divided our investigations into two main parts; an interface fluctuation 
part and a second part where the colloid fluctuations are included.
In the interface fluctuation part, the positions of the colloids and thus the contact lines are fixed 
 and therefore the problem reduces to the ``usual" Casimir problem with Dirichlet boundary
conditions. The Casimir interaction is determined by multipoles of an auxiliary field on the 
(1d) contact lines
which is similar to the source field in the language of Ref.~\cite{jaffe1}. 
In the second part, we include that
the colloid position may fluctuate in all possible ways (height and tilts)
which turns out to lead to Casimir interactions determined by multipoles of an 
auxiliary field defined on the 2d domain enclosed by the contact lines. 
The fluctuating boundaries result in certain restrictions on these multipoles.
(Incidentally, we note that for small fluctuations 
amplitudes, an exact separation of interface and colloid fluctuation in the partition sum
is possible \cite{martin1, martin2}. This route, however, appears to be very difficult to follow
for other than spherical colloids and is not taken here.)} 

{The paper is structured as follows.
In Sec.~\ref{sec:model}, we introduce the effective
Hamiltonian of the model system and the partition function 
which is defined by functional integrals over colloid and interface fluctuations. 
The implementation of the different boundary conditions is discussed.
In Secs.~\ref{sec:intflucpart} and \ref{numerics},
we determine the fluctuation--induced force 
for the interface fluctuation part in the long and short distance regimes analytically and 
for intermediate distances numerically, respectively . 
In Sec. \ref{sec:Kardar}, we include the colloid fluctuations to the problem 
and discuss the modified long--distance asymptotics of the Casimir interaction.
Sec.~\ref{summary} contains a brief summary. Some technical details of the calculations
have been relegated to Apps.~\ref{elliptic_coordinates}--\ref{app:se}. }
%
%
\section{Model}\label{sec:model}
The investigated system consists of two nano- or microscopic, uncharged spheroidal colloids
with principal axes $a$, $b$, $b$ ($a>b$), 
which are trapped at the interface of two fluid phases I and II. The effective interaction between the
colloids is mediated by thermal height fluctuations of the (sharp) interface. 
{Without fluctuations, the equilibrium interface is flat and is set to be at $z = 0$.
The corresponding equilibrium position of the colloids is assumed to be symmetrical with respect to $z\to -z$,
such that at the contact line the contact angle is $\pi/2$.}
The elliptic cross-section of the ellipsoids with the equilibrium interface is denoted by $S_{i,{\rm ref}}$ 
which is an ellipse with major and minor axes $a, b$, respectively. $S_{i,{\rm ref}}$ may also be expressed 
in confocal elliptic coordinates by the elliptic radius $\xi_0$, see App.~\ref{elliptic_coordinates}
for the coordinate definitions. The equilibrium interface at $z = 0$ without the two elliptic holes
$S_{i,{\rm ref}}$ cut out by the colloids is termed the reference meniscus  
$S_{\rm men,ref}=\mathbb{R}^2\setminus \cup_iS_{i,\rm ref}$. Deviations from this planar reference meniscus 
are considered to be small, without overhangs and bubbles, therefore the Monge representation 
$(x,y,z=u(x,y))=({\bf x},z=u({\bf x}))$ is employed to describe the interface position. 
The colloids are of Janus type, thus the contact line is always pinned to their surface. 
The total Hamiltonian of the system which is used for calculating the free energy costs of thermal fluctuations around the flat interface is determined by the change in interfacial energy of the interface I/II:
\bea\label{ham2}
\mc{H}_{\rm tot}=\gamma\Delta A_{\rm men}
&=&
\gamma \int_{S_{\rm men}} d^2x\,\sqrt{1+ (\nabla u)^2}
-\gamma \int_{S_{\rm men,ref}} d^2x
\nonumber\\
&\approx&
\frac{\gamma}{2} \int_{S_{\rm men,ref}} d^2x\,  (\nabla u)^2
+\gamma \Delta A_{\rm proj} \,.
\eea 
\begin{figure}
 \includegraphics[scale=0.5]{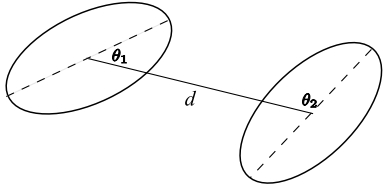}
\caption{Top view on the system}
\label{plane}
\end{figure}
Where $\gamma\Delta A_{\rm men}$ expresses the energy needed for creating the additional meniscus area associated with the height fluctuations. In Eq.~(\ref{ham2}), $S_{\rm men}$ is the meniscus area projected onto the plane $z=0$
(where the reference interface is located)
and $S_{\rm men,ref}$ is the meniscus in the reference configuration mentioned
above. 
{The first line of Eq.~(\ref{ham2})
constitutes the  drumhead model which is well-known in the renormalization group
analysis of interface problems, but is also used for 
the description of elastic surfaces 
(c.f. Ref.~\cite{Jas86}).
In the second line we have applied a small gradient expansion
which is valid for slopes $|\nabla u|\ll 1$ and which provides the long wavelength
description of the interface fluctuations we are interested in. 
The small gradient expansion entails that
\begin{equation}
 \label{aproj}
\Delta A_{\rm proj}=  \int_{S_{\rm men}\setminus
S_{\rm men,ref}}\!\!\! d^2x\,\sqrt{1+(\nabla u)^2} \approx
\int_{S_{\rm men}\setminus 
S_{\rm men,ref}}\!\!\!d^2x
\end{equation}
 is the change
in projected meniscus area with respect to the reference configuration.
We rewrite this change in projected meniscus area in terms of the interface position 
$f_i=u(\partial S_{i,{\rm ref}})$ at the reference contact line ellipses $\partial S_{i,{\rm ref}}$.
$f_i$ corresponds (in second order approximation) to the contact line of
the colloid $i$ with fluctuating center position $h_i$ and fluctuating orientation.
The contact line which is a function of the elliptic angle $\eta$ only { (see App.~\ref{elliptic_coordinates} for the definition of elliptic coordinates)} is expanded
as
}
\be\label{conlin}
f_i = 
u(\partial S_{i,\rm ref})=\sum_{m=0}  (P_{im}\,\cos (m\eta_i) + Q_{im}\,\sin (m\eta_i))
\;
\ee
and we refer to 
the coefficients $P_{im}$ and $Q_{im}$ as 
boundary multipole
moments below.
{
The desired expression of $\Delta A_{\rm proj}$ in terms of boundary multipole moments
proceeds as discussed in Ref.~\cite{Oet05} and allows us to identify it as a sum over
boundary Hamiltonians ${\cal H}_{i,\rm b}$ for each colloid $i$ (see also App.~\ref{app:Hb}):
}
\begin{eqnarray}
\label{Hb}
\gamma \Delta A_{\rm proj}
&\equiv&
\sum_i\mc{H}_{{\rm b},i}[f_i]
\nonumber\\
&=&
\sum_i\frac{\pi\gamma}{2}\left( \tanh \xi_0\, P_{i1}^2 + \coth \xi_0\, Q_{i1}^2 \right)\;.
\end{eqnarray}
Putting Eqs.~(\ref{ham2}) and (\ref{Hb}) together, 
the total change in interfacial energy is the sum 
\bea\label{Htot2}
\mc{H}_{\rm tot}
&=&
\mc{H}_{\rm cw}+\mc{H}_{\rm b,1}+\mc{H}_{\rm b,2}
=
\frac{\gamma}{2} \int_{S_{\rm men,ref}} d^2x\,  (\nabla u)^2
+\mc{H}_{\rm b,1}+\mc{H}_{\rm b,2}
\eea
of the capillary wave Hamiltonian $\mc{H}_{\rm cw}$
which describes the energy differences
associated with the additional interfacial area over the reference configuration 
and the boundary Hamiltonians $\mc{H}_{\rm b,i}$
{which can be viewed as the energy cost due to fluctuations of the contact line
(and which in turn are caused by colloid height and tilt fluctuations).}
As is well-known, the
 Hamiltonian $\mc{H}_{\rm cw}$ is plagued with
both a short-wavelength and a long-wavelength
divergence which, however,  can be treated by physical cutoffs.
The natural
short-wavelength cut-off is set by
the molecular length-scale $\sigma$ of the
fluid at which the capillary wave model
ceases to remain valid. 
The long wavelength divergence is reminiscent to the fact
that the capillary waves are Goldstone modes. 
Of course, in real systems 
the gravitational field provides a natural damping for capillary waves.
Accounting also for the costs  in gravitational energy associated 
with the interface height fluctuations,
therefore, introduces a long wavelength cutoff
and leads to an additional term (``mass term'') in the capillary wave Hamiltonian,
\bea\label{capwav}
\mc{H}_{\rm cw}
&=&
\frac{\gamma}{2} \int_{S_{\rm men,ref}} d^2x\,  
\left[(\nabla u)^2+\frac{u^2}{\lambda_c^2}
\right]
\eea
 Here the capillary length is given by 
 $\lambda_c = [\gamma/(|\rho_{\rm II}-\rho_{\rm I} |\, g)]^{1/2}$, where $\rho_i$
 is the mass density in phase $i$ and $g$ is the gravitational constant. 
 Usually, in simple liquids, $\lambda_c$ is in the range of millimeters
and, therefore, is by far the longest length scale in the
system. In fact, here it plays the role of a long wavelength cutoff
of the capillary wave Hamiltonian $\mc{H}_{\rm cw}$, 
and we will discuss our results in the limit  $\lambda_c \gg R$ 
and $\lambda_c \gg d$.
However, as we will see below,
care is required when taking
 the limit $\lambda_c \to \infty$ (corresponding to $g \to 0$), since 
logarithmic divergencies appear \cite{Saf94}.
Another common way to introduce a long-wavelength cut-off
is the finite size $L$ of any real system.
As discussed in Ref.~\cite{Oet05}, the precise way of incorporating
the long-wavelength cut-off is unimportant for the effects on the colloidal
length scale. As an example, 
in both approaches the
width of the interface related to
the capillary wave is logarithmically divergent,
$\langle u(0)^2\rangle \sim \ln \lambda_c[L]/\sigma $.
%
\begin{figure}
\includegraphics[width=0.6\textwidth]{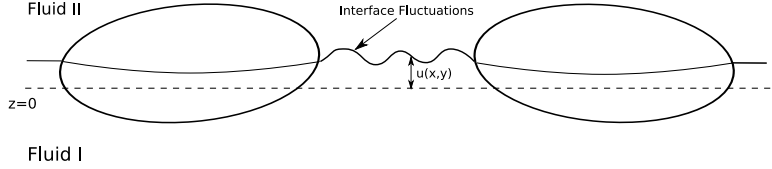}
\caption{Side view on the system}
\label{fig1}
\end{figure}
Via the integration
domain of $\mc{H}_{\rm cw}$, 
the total Hamiltonian  of the system, Eq.~(\ref{Htot2}), implicitly  depends
on the geometric configuration.
{ This leads to a free energy ${\cal F}(d,\theta_1,\theta_2)$ which depends on the
distance $d$ between the colloid centers
and the orientation angles $\theta_1$ and $\theta_2$ of their major axes with respect to
the distance vector joining the colloid centers (see Fig.~\ref{plane}). 
The free energy is related to the partition function ${\mathcal Z}(d,\theta_1,\theta_2)$ of the system
by}
\bea
 {\mathcal F}(d,\theta_1,\theta_2)= -k_{\rm B} T \ln {\mathcal Z}(d,\theta_1,\theta_2)
\eea
The partition function
is obtained by a functional integral over all possible interface 
configurations $u$ and  boundary configurations $f_i$; the relation between interface and boundary 
configurations is included by 
$\delta$-function constraints,  
\begin{equation}\label{Z1}
\mathcal{Z}=\mathcal{Z}_0^{-1} \int \mathcal{D}u\,
\exp \left\{-\frac{\mathcal{H}_{\rm cw}[u] }{k_{\rm B}T} \right\}\;
\prod_{\rm i=1}^2 
\int \mathcal{D}f_i
\prod_{{\bf x}_i \in \partial S_{i,\rm ref}}
\delta [u({\bf x}_{ i})-f_{ i}({\bf x}_{ i})]
\exp \left\{-\frac{\mathcal{H}_{{\rm b},i}[f_i]}{k_{\rm B}T} \right\}
\;.
\end{equation}
Here $\mathcal{Z}_0$ is a normalization factor such that
$\mathcal{Z}(d\to\infty)=1$
and ensures a proper regularization of
the functional integral.
Via the $\delta$-functions the interface field $u$ is
coupled to the contact line height $f_i$ and therefore,
the boundary Hamiltonians $\mc{H}_{i,\rm b}$
have a crucial influence on the resulting effective
interaction between the colloids.\newline
{The kind of possible contact line fluctuations $f_i$ is solely  determined
by the colloid fluctuations since the contact line is pinned. These fluctuations
are vertical
fluctuation of the colloids on the axis normal to the equilibrium interface (height)
and orientational fluctuations around that axis (tilts).}
In order to incorporate various boundary counditions into the solutions,
we categorize them into three cases:
\begin{itemize}
\item[(A)] colloids are fixed in the reference configuration, thus there are no
integrations over the boundary terms. 
\item[(B)] colloid heights 
fluctuate freely without tilting, thus the boundary monopoles
 must be included in the integration measure 
 so that {$\mc{D}f_i= dP_{i0}$}.
\item[(C)] unconstrained height and tilt fluctuations. 
Up to second order in the tilts this corresponds
to the inclusion of boundary dipoles in the integration measure, thus
{$\mc{D}f_i=dP_{i0}\, dP_{i1}\, dQ_{i1}$}.
\end{itemize}
{Case (A) corresponds to the ``standard" Casimir effect in 2d with Dirichlet
boundary conditions $f_i=u(\partial S_{i,{\rm ref}})=0$. 
We call this the interface fluctuation part and it will be treated in Sec.~\ref{sec:intflucpart}.
Note that a short summary of this part has already been given in Ref.~\cite{meandmartin}.
The inclusion of the colloid height and tilt fluctuations in (B) and (C) is given in 
Sec.~\ref{sec:Kardar}.}
\section{Interface fluctuation part }\label{sec:intflucpart}
The partition function  $\mathcal{Z}_{\rm in}$ for fixed contact lines $f_i=0$ is given by
\begin{equation}\label{Zsep}
 \mathcal{Z}_{\rm in} = \mathcal{Z}_0^{-1} \int \mathcal{D}u\,
\prod_{\rm i=1}^2 \prod_{{\bf x}_i \in \partial S_{i,{\rm ref}}}\delta ( u({\bf x}_i))
\exp \left\{-\frac{\mathcal{H}_{\rm cw}[u]}{k_{\rm B}T} \right\}\;.
\end{equation}
The disapearance of the interface fluctuations
 at the colloids boundaries is included by 
the Dirac delta function.
In this section, analytical expressions for the
fluctuation induced force
in the intermediate asymptotic regime $a\ll d \ll \lambda_c$
are calculated.
We express the $\delta$-functions 
in Eq.~(\ref{Zsep})
by  their integral representation via
 auxiliary fields $\psi_i ({\bf x}_i)$ defined on the
{reference contact lines} $\partial S_{i,\rm ref}$. This enables us to integrate out 
the field $u$ leading to
\begin{eqnarray}
\label{Zaux}
 \mathcal{Z}_{\rm in} =
{\mathcal{Z}_0^{\prime}}^{-1} \int \prod_{i=1}^2 \mc{D}\psi_i\,
\exp\left\{
-\frac{k_{\rm B}T}{2\gamma}\sum_{i,j=1}^2
\int_{\partial S_{i,\rm ref}}d\ell_i \int_{\partial S_{j,\rm ref}}d\ell_j\,
\psi_i ({\bf x}_i)\,G(|{\bf x}_i-{\bf x}_j|)\,\psi_j({\bf x}_j)\right\}\;,
\end{eqnarray} 
where $d\ell_i$ is the infinitesimal line segment
 on the circles $\partial S_{i,\rm ref}$.
{After this integration, the normalization factor is changed, 
${\mathcal Z}_{0} \to {\mathcal Z}_{0}'$, such that still 
${\mathcal Z}_{\rm in}(d\rightarrow \infty)=1$ holds}.
In Eq.~(\ref{Zaux}) we introduced the Greens function of the operator 
$(-\Delta + \lambda_c^{-2})$
which is given by
$G({\bf x})=K_0(|{\bf x}|/\lambda_{ c})/(2\pi)$ where $K_0$ is the modified Bessel
function of the second kind.
 In the range $d/\lambda_c \ll 1$ and $r_0/\lambda_c \ll 1$,
we can use 
 the asymptotic form of the $K_0$ for small arguments, such that
$2\pi\,G(|{\bf x}|) \approx -\ln(\gamma_{\rm e} |{\bf x}|/\,2\lambda_c) $. Here,
$\gamma_{\rm e} \approx 1.781972$  is the Euler-Mascheroni constant exponentiated. 
We introduce auxiliary multipole moments as the
Fourier-transforms of 
the auxiliary fields $\psi_i$ on the reference contact line $\partial S_{i,\rm ref}$,
\bea
\wh{\psi}_{im}^c
&=&
\int_0^{2\pi}
d\eta_i\,h (\eta_i) \, {\cos(m\eta_i)} \,\psi_{i}({\bf x}_i(\eta_i))\;, \nonumber\\
\label{auxmomdef}
\wh{\psi}_{im}^s
&=&
\int_0^{2\pi}
d\eta_i\,h (\eta_i) \, {\sin(m\eta_i)} \,\psi_{i}({\bf x}_i(\eta_i)) \;,
\eea
{where $\eta_i$ is the elliptic angle pertaining to a coordinate system centered
around each colloid $i$, respectively, such that the $x$--axis in this colloid--specific
coordinate system joins the two foci of $S_{i,{\rm ref}}$.}
Furthermore, $h(\eta_i)$ is the scale factor in elliptic coordinates
(see App.~\ref{elliptic_coordinates}). 
{The lengthy calculation leading to the multipole {(Fourier)} decomposition for the Greens function
$G(|{\bf x}_i-{\bf x}_j|)$ (for general orientations $\theta_1$ and $\theta_2$ of the ellipsoids) is
given in App. \ref{app:schwing}. The final results is collected in
Eq.~(\ref{gself}) and Eqs.~(\ref{final_exp})--(\ref{c_llmn}).}
Using this,  the double integral in the
exponent of Eq.~(\ref{Zaux}) can be written as a double sum
over {the auxiliary multipole moments}  (Fourier components), 
consisting of a self-energy part $G_{\rm self}$ when
$x_i$ and $x_j$ reside on one ellipse and $G_{\rm int}$ when the points
 $x_i$ and $x_j$ reside on different ellipses, respectively.
The functional integral over the auxiliary fields becomes a product
of integrals over their multipole moments,
$\mc{D}\psi_i = d\wh{\psi}_{i0}\prod_{j=1}^\infty d\wh{\psi}^c_{ij}d\wh{\psi}^s_{ij}$, and the resulting
partition function then reads
\begin{eqnarray}\label{Zauxmom}
  \mc{Z}_{\rm in}
  &=&
  {\mathcal{Z}_0^{\prime}}^{-1}\int \prod_{i=1}^2 \mc{D}\psi_i\,
  \exp\left\{-
    \frac{k_{\rm B}T}{2\gamma}
    \left(
      \begin{array}{c}
        {\bf\widehat{\Psi}}_1\\
        {\bf\widehat{\Psi}}_2
      \end{array}
    \right)
    ^{\rm T}
    \left(\begin{array}{cc}
        \wh{ \bf G}_{\rm self} & \wh{\bf{ G}}_{\rm int} \\
        \wh{\bf{G}}_{\rm int} & \wh{\bf{G} }_{\rm self}
      \end{array}\right)
    \left(
      \begin{array}{c}
        {\bf \widehat{\Psi}}_1\\
        {\bf \widehat{\Psi}}_2
 \end{array}
\right)
\right\}\;,
\end{eqnarray}
where the vectors 
${\bf  \widehat{\Psi}}_i= ({\bf\widehat{\Psi}}_i^c,{\bf \widehat{\Psi}}_i^s)$ with 
${\bf\widehat{\Psi}}_i^c=(\wh{\psi}_{i0}^c,\wh{\psi}^c_{i1},\dots)$ and 
${\bf\widehat{\Psi}}_i^s=(\wh{\psi}^s_{i1},\wh{\psi}^s_{i2},\dots)$
contain the auxiliary multipole moments of colloid $i$. 
The coupling matrix ${\bf \wh{G}}$ which contains the 
Fourier modes of the Greens function $G({\bf x}_i-{\bf x}_j)$
has a block structure. The self energy
submatrix $\bf{\wh{G}}_{\rm self}$ which describes
the coupling between auxiliary moments of the same colloid
are diagonal, and its form can be determined from definition (\ref{auxmomdef}) and Eq.~(\ref{gself}). 
\begin{equation}
 \label{self-elements}
2\pi\,({\bf  \widehat{\Psi}}_i)^{\rm T} \wh{\bf{G} }_{\rm self} {\bf  \widehat{\Psi}}_i 
=
-\ln{\frac{\gamma_{\rm e} {a^{\prime} e^{\xi_0}}}{8\lambda_c}} {({\wh\psi_{i0}^c})^2}
+ 2\sum_{n=1}
\frac{e^{-n \xi_0}}{n} \left[ \cosh (n\xi_0) \, ({\wh\psi_{in}^c})^2 + \sinh (n\xi_0) \,  ({\wh\psi_{in}^s})^2\right]  \;,
\end{equation}
%
 {where ${a^{\prime}}^2 = a^2 - b^2$}. The offdiagonal blocks $\bf{\wh{G}}_{\rm int}$ characterise
the interaction between the multipole moments residing on different
colloids. It is convenient to split the matrix into a block structure describing the interaction
of cosine and sine multipoles: 
\begin{eqnarray}\label{gint}
 2\pi\,({\wh{\Psi}_1})^{\rm T}\, \wh{\bf{G} }_{\rm int}\, \wh{\Psi}_2
  &=&
     \left(
      \begin{array}{c}
        {\bf\widehat{\Psi}}_1^c\\
        {\bf\widehat{\Psi}}_1^s
      \end{array}
    \right)
    ^{\rm T}
    \left(\begin{array}{cc}
        \wh{ \bf G}_{\rm int}^{cc} & \wh{\bf{G}}_{\rm int}^{sc} \\
        \wh{\bf{G}}_{\rm int}^{sc} & \wh{\bf{G} }_{\rm int}^{ss}
      \end{array}\right)
    \left(
      \begin{array}{c}
        {\bf \widehat{\Psi}}_2^c\\
        {\bf \widehat{\Psi}}_2^s
 \end{array}
\right)
\end{eqnarray}
The matrix elements of the such defined submatrices follow from
Eqs.~(\ref{final_exp})--(\ref{c_llmn}), and are explicitly given by:
\begin{eqnarray}
\label{int-cc00}
 \left(\wh{\bf{G} }_{\rm int}^{cc}\right)_{0\,0} & = &
-\ln\left( \frac{\gamma_{\rm e} d}{2\lambda_c}\right)
\end{eqnarray}
\begin{eqnarray}
\label{int-ccmn}
 \left(\wh{\bf{G} }_{\rm int}^{cc}\right)_{mn} & = &
\sum_{l=0}\left(\frac{a^{\prime}}{4d}\right)^{m+n+2l} A_{mnl}^c(\theta_1,\theta_2)\cosh (m\xi_0) \, \cosh (n\xi_0) 
\end{eqnarray}
\begin{equation}
\label{int-ss}
 \left(\wh{\bf{G} }_{\rm int}^{ss}\right)_{mn}  = 
-\sum_{l=0}\left(\frac{a^{\prime}}{4d}\right)^{m+n+2l} A_{mnl}^c(\theta_1,\theta_2)\sinh (m\xi_0) \, \sinh (n\xi_0) 
\end{equation}
\begin{eqnarray}
 \label{sc}
\left(\wh{\bf{G} }_{\rm int}^{sc}\right)_{mn} & = &
  \sum_{l=0}\left(\frac{a^{\prime}}{4d}\right)^{m+n+2l} A_{mnl}^s(\theta_1,\theta_2) \sinh (m\xi_0) \, \cosh (n\xi_0)
\end{eqnarray}
From Eq.~(\ref{Zauxmom}) we find that
the fluctuation part of the free energy reads
\bea\label{Fflucaux}
\mc{F}_{\rm in}
=-k_{\rm B}T\ln {\mathcal Z}_{\rm in}
=
-\frac{k_{\rm B}T}{2}\ln( \det{\widehat{\bf G}}) + {\rm const.} \;,
\eea
{where ${\rm const.} = -k_{\rm B}T \ln {\mathcal{Z}_{\rm 0}}^{\prime}$.}
{The factors $A_{mnl}^{c[s]}(\theta_1,\theta_2)$, given in Eqs.~(\ref{a_mnl}) and (\ref{c_llmn}), 
contain the dependence on the orientation angles $\theta_1$ and $\theta_2$
of the ellipsoids (see Fig.~\ref{plane}). As can be seen from above, the interaction
coefficients  $\left(\wh{\bf{G} }_{\rm int}^{c[s]c[s]}\right)_{mn}$ between multipoles of order
$m$ and $n$ take the form of a series in $1/d$, starting at $1/d^{m+n}$. (For spherical colloids,
this multipole interaction coefficient only contains the order $1/d^{m+n}$ \cite{martin2}.)}
In principle, the matrix $\wh{\bf G}$ is  infinite dimensional and
{$\det({\widehat{\bf G}})$ is divergent and its regularisation is provided by the normalization factor ${\mathcal{Z}_0^{\prime}}$.}
The explicit series for the elements of $\bf{\wh{G}}_{\rm int}$
allows for a systematic expansion of
the logarithm in Eq.~(\ref{Fflucaux}) 
in powers of $1/d$,
\bea\label{Fflucexp}
\mc{F}_{\rm in}(d)
&=&
k_{\rm B}T
\sum_{n}f_{2n}^{\rm in}
\,\left(\frac{1}{d}\right)^{2n}\;,
\eea
where the coefficients $f_{2n}^{\rm in}$ depend 
on the logarithms
$-\ln(\gamma_{\rm  e}d/\,2\lambda_c))$ 
and $-\ln(\gamma_{\rm e}a^{\prime}e^{\xi_0})/\,8\lambda_c)$,
{as well as the angles $\theta_1$ and $\theta_2$}.
 The number of auxiliary multipoles included in
the calculation of the asymptotic form of
$\mc{F}_{\rm fluc}$ in Eq.~(\ref{Fflucexp})
is determined by the desired order in $1/d$. 
{Inclusion of multipoles up to order $n$ leads to an asymptotics correct up to 
$1/d^{2n}$.}
In the limit $\lambda_c/d \to \infty$ 
the free energy expansion coefficients in Eq.~(\ref{Fflucexp})
 up to fourth order are\footnote{{Note that a factor of $1/(2\ln[2d/(a+b)])$ is missing in the expression
for $f_2^{\rm in}$ in Eq.~(29) of Ref.~\cite{meandmartin}.}}
\begin{eqnarray}
 f_{0}^{\rm in} &=& \frac{1}{2}\ln\ln\left(\frac{4d}{a+b}\right)+{\rm const.} \nonumber \\
 \label{ares}
 f_2^{\rm in} &=&-\frac{1}{2\ln\left(\frac{\di 4d}{\di a+b}\right)}\left[\frac{(a+b)^2}{16}+ \frac{3}{32} (a^2-b^2)(\cos(2\theta_1)+\cos(2\theta_2))\right] \\
f_4^{\rm in}  &=&  -\frac{1}{2^{11} } \left\lbrace\frac{1}{\ln\left(\frac{\di 4d}{\di a+b}\right)}
\left[ \phantom{\frac{1}{1}} 16(a-b)(a+b)^3(\cos(2\theta_1) + \cos(2\theta_2)) \right. \right. \nonumber\\
&& + 11(a^2-b^2)^2 (\cos(4\theta_1) + \cos(4\theta_2)) \nonumber\\
&& \left.+ 44(a^2-b^2)^2 \cos(2\theta_1+2\theta_2) + 6(a+b)^4\phantom{\frac{1}{1}}\right]  \nonumber\\
&&+\left. [ 8(a^2-b^2)^2\cos(2\theta_1+2\theta_2)+8(a+b)^4 ] \phantom{\frac{1}{\frac{\di 1}{\di 1}}}\right\rbrace -\frac{1}{2}\left(f_2^{\rm in}\right)^2\  \nonumber
\end{eqnarray}
{The double--logarithmic divergence in $d$ in the leading coefficent $f_{0}^{\rm in}$
is a reflection of the fact that the interface itself becomes ill--defined for
$\lambda_c\to\infty$ due to the capillary waves. For the Casimir force itself, however,
we find a finite value for all $d$ in the limit $\lambda_c\to \infty$.} 
{Anisotropies in the Casimir interaction appear here first in the subleading term
$f_2^{\rm in}$. Their angular dependence stems from the monopole--dipole interaction of
the auxiliary field, and the attraction is maximal if both ellipses are aligned tip--to--tip.}

 In the
 opposite limit of small surface--to--surface distance $h=d-d_{cl} \ll d_{cl}$, 
(where $d_{cl}$ is the distance of the closest approach between ellipses)  the
 fluctuation force can be calculated by using
the Derjaguin (or proximity) approximation  \cite{Der34}.
 It consists in replacing the local
 force density on the contact lines by the result for the fluctuation force
 per length $f_{\rm 2d}(\tilde h)$ between two parallel lines with
 a separation distance ${\tilde h}$ and integrating over the
 two opposite contact lines to obtain the total effective force
 between the colloids.
The Casimir force density between two parallel
surfaces was calculated in Ref.~\cite{Li91}
in a general approach for arbitrary dimensions.
Applied to two dimensions
we obtain the  force line density  $f_{\rm 2d}(\tilde h)= -k_{\rm B}T\,\pi/(24 {\tilde h}^2)$. 
{Integrating this density over the opposing contact lines
yields \cite{meandmartin}
 \begin{eqnarray}
 \label{fflucderja}
  F_{\rm in} \approx  
  -\frac{\pi k_{\rm B}T}{24}\int_{-\infty}^{+\infty}dy\,
 \frac{1}{\left(h+\frac{y^2}{2}(\frac{1}{R_1}+\frac{1}{R_2})\right)^2}
  = -k_{\rm B}T\, \frac{\pi^2}{48h^{\frac{3}{2}}}\sqrt{\frac{2}{\frac{1}{R_1}+\frac{1}{R_2}}}\,
  +\mc{O}(h^{-1/2})
   \;.
  \end{eqnarray}
Here, $R_1$ and $R_2$ are the curvature radii of the two ellipses at
the end points of the distance vector of the closest approach.
It is seen that the fluctuation force diverges as $h^{-3/2}$
upon contact of the ellipsoids ($h\to 0$).
}
 
\subsection{Intermediate distances: Numerical calculation}\label{numerics}
\begin{figure}
\begin{tabular}{cc}
\includegraphics[width=0.45\textwidth]{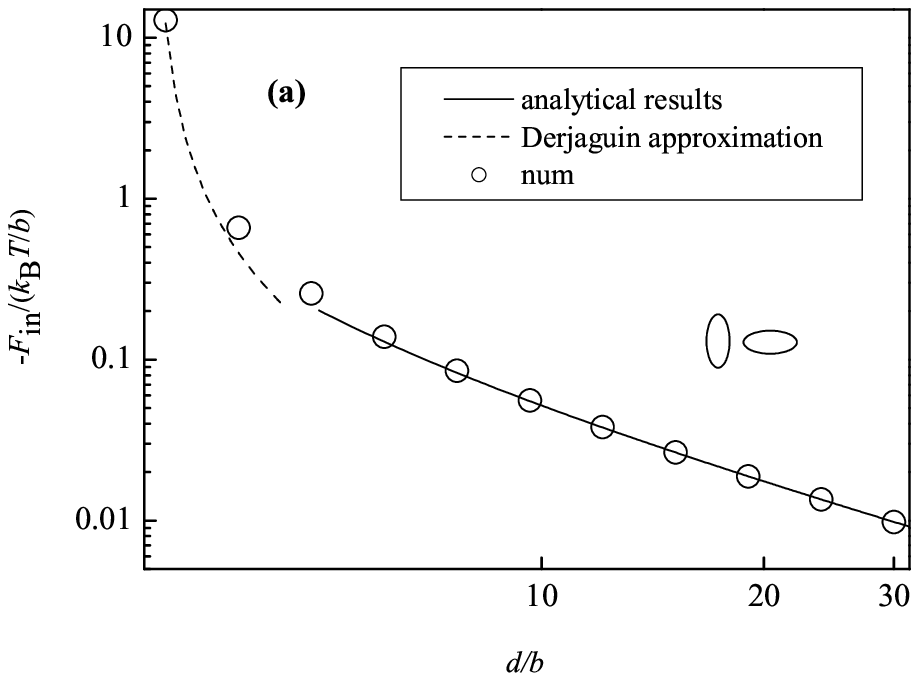}
\hspace*{0.5cm}
&
\includegraphics[width=0.45\textwidth]{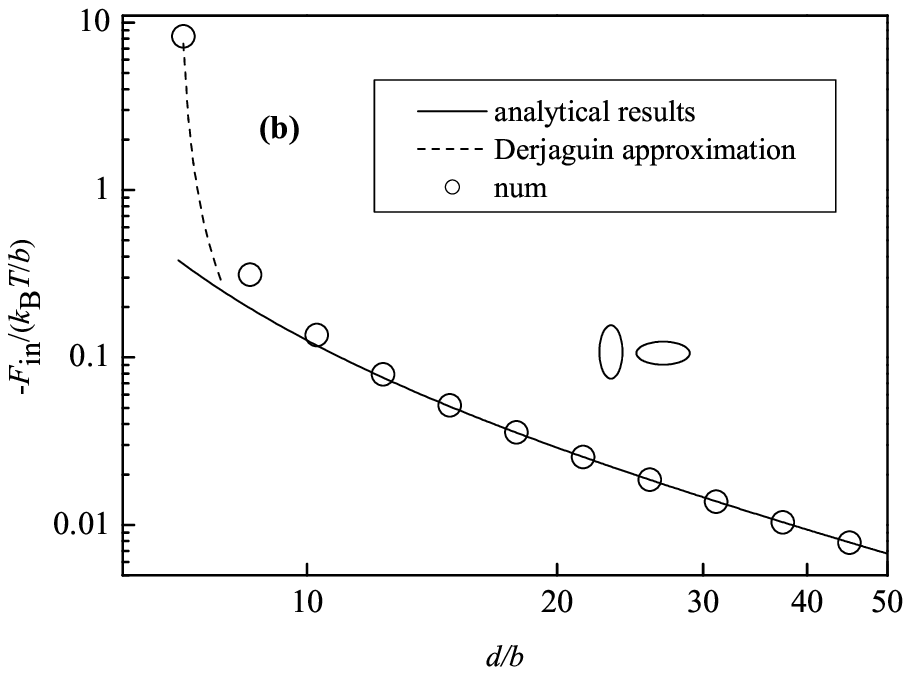}
\vspace*{0.05cm}
\\
\includegraphics[width=0.45\textwidth]{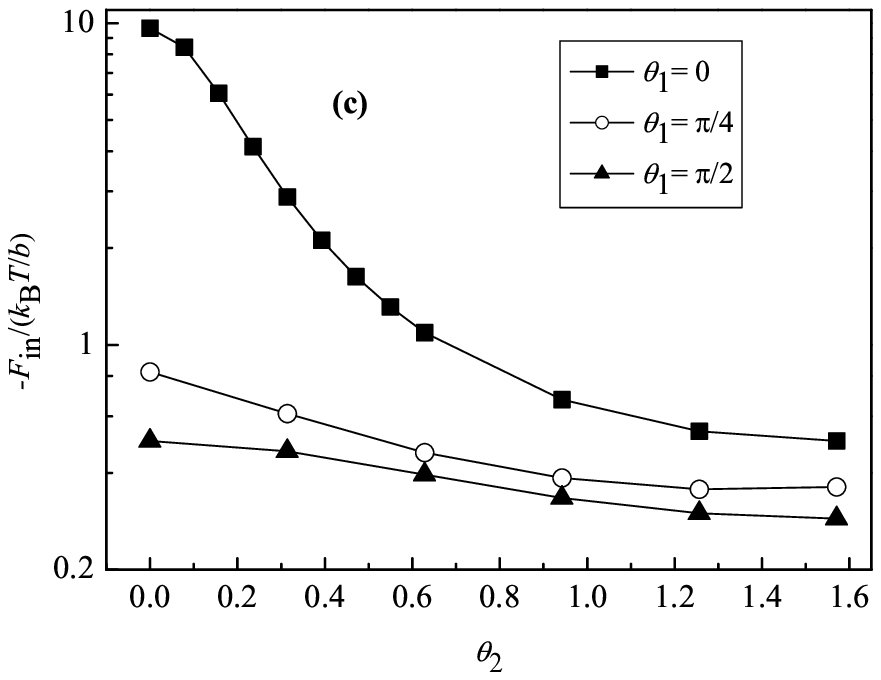}
\hspace*{0.5cm}
&
\includegraphics[width=0.45\textwidth]{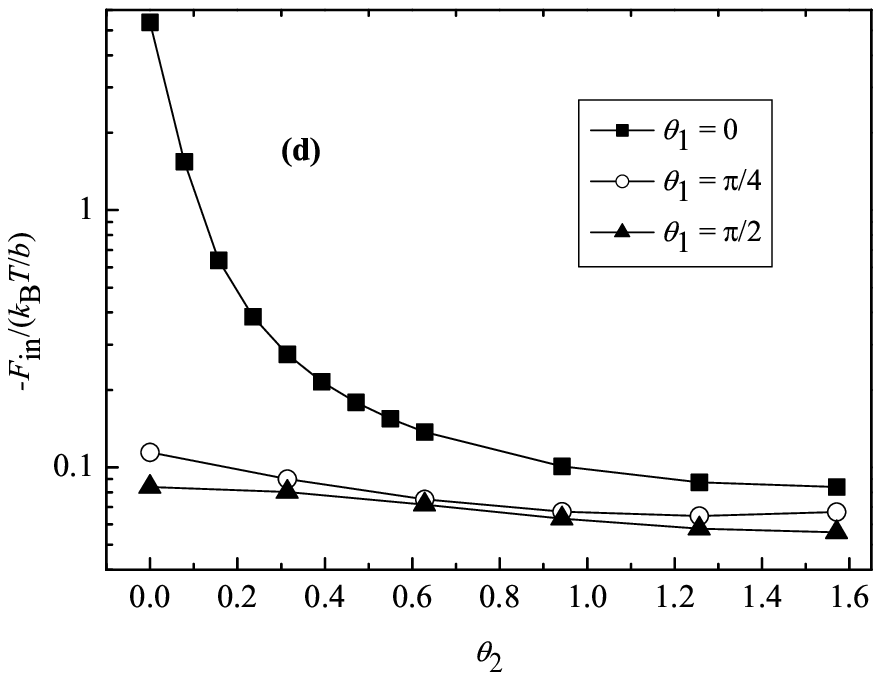}
\end{tabular}
\caption{{\bf(a,b)}Comparison of the numerical results for the interface fluctuation 
Casimir force (symbols) with the analytical expressions
in the asymptotic ranges of large colloid separations $d \gg a,b$
(full line) and small
surface-to-surface distance $h=d-d_{cl} \ll b$ (dashed line)
{\bf (c,d)} numerical Casimir interaction between two fixed ellipsoids
 trapped at the interface as a function of their orientation, for $d=4.1$ and $d=12.1$, respectively.
}
\label{figures}
\end{figure}

For intermediate distances {$d-d_{cl} \simeq a$} the fluctuation induced
force has to be calculated numerically.
{This can be done in principle by including a number of multipoles in the numeric evaluation
of the determinant in Eq.~(\ref{Fflucaux}), see Ref.~\cite{jaffe1}. 
In order to avoid the algebraic evaluation of the multipole coefficients of $\bf{\hat G}$, it is possible 
to apply a method which was introduced in Ref.~\cite{Bue04}.}
The starting point is Eq.~(\ref{Zaux}) for the
partition function $Z_{\rm in}$.
{Introducing a mesh  with $N$ points
$\eta_{ij}$, $j =1\dots  N$,
on the reference contact line
$\partial S_{i,\rm ref}$ converts 
the double integral in the exponent
to a double sum. Thus the functional integrals
over the auxiliary fields $\psi_i(\eta_i)$ are replaced by ordinary
Gaussian integrals over the $\psi_i({\bf x}_i(\eta_{ij}))$,
$\mc{D}\psi_i\simeq \prod_{j=0}^Nd\psi_i({\bf x}_i(\eta_{ij}))$.
In the exponent, the $\psi_i({\bf x}_i(\varphi_{ij}))$ are
coupled by a matrix $\bf{G}$ with
elements $G_{ii'}^{jj'}=G(|{\bf x}_i(\eta_{ij})-{\bf x}_{i'}(\eta_{i'j'})|)$.}
performing the Gaussian integrals and 
disregarding divergent and $d$-independent terms immediately
leads to 
$\mc{F}_{\rm in}=(k_{\rm B}T/2)\ln\det({\bf G}_\infty^{-1}{\bf G}(d))$
for the fluctuation free energy.
Here, ${\bf G}_\infty\equiv \lim_{d\to\infty}{\bf G}(d)$.
It contains the self energy contributions and is
needed for the regularization of the free energy.
Deriving with respect to $d$, the Casimir
force can be written as 
\be\label{numforce}
F_{\rm in}(d)
=-\frac{k_{\rm B}T}{2}
\,{\rm tr}\left[
{\bf G}(d)^{-1}\partial_d{\bf G}(d)
\right]\;.
\ee
The advantage of the direct calculation of the
force is that Eq.~(\ref{numforce}) does not contain any 
divergent parts which would require regularization,
thus easing the
numerical treatment considerably.
The determinant is computed by
using a standard LU decomposition \cite{press02}.
We find  good 
convergence of the numerical routine.
{ The convergence can be sped up by distributing more points in the regions where the 
ellipses face each other. We note that computing the force by the multipole series 
seems to be more efficient \cite{jaffe1}; this can partially be compensated by the point 
distribution on the ellipses.}
{In Fig.~\ref{figures}a (ellipse aspect ratio 2) and \ref{figures}b (aspect ratio 6) 
we compare the analytical results of Eqs.~(\ref{Fflucexp}) 
and the Derjaguin approximation (Eq.~(\ref{fflucderja}) with the numerical results. 
As it is shown the analytical expressions show very good agreement with the numerical data points 
for both long- and short range behavior and almost cover the whole distance regime.
At large distances $d$, the leading term of the free energy expansion
 in Eqs.~(\ref{Fflucexp}) mainly determines the behavior of 
the Casimir interaction because of its long-ranged nature,
hence the orientation dependence of the subleading terms can be neglected.
In order to demonstrate the anisotropy of the Casimir interaction, we show
results for a fixed, intermediate distance $d$ between ellipsoid centers and varying
orientation $\theta_2$ of the second ellipsoid, see Fig.~\ref{figures}c (aspect ratio 2, $d/b=4.1$)
and \ref{figures}d (aspect ratio 6, $d/b=12.1$).
The orientation of the first ellipse was fixed to three values,
$\theta_1=0$, $\theta_1=\pi/4$ and $\theta_1=\pi/2$. 
As can be seen, the fluctuation--induced interaction is maximally attractive for $\theta_1= \theta_2=0$ 
(tip--to--tip configuration). When $\theta_2$ deviates from zero 
then the resulting force reduces. This behavior holds for both aspect ratios 2 and 6.}

\section{Inclusion of colloid fluctuations}\label{sec:Kardar}
{In general, the inclusion of colloid height and tilt fluctuations into the partition function
(Eq.~(\ref{Z1})) can be realized by an approach used in Refs.~\cite{martin1,martin2}. 
In this approach,
 the partition function is split into a product of a colloid fluctuation part and the
interface fluctuation part.
The latter contains only the contribution of the fluctuating interface, with Dirichlet 
boundary conditions on the colloid surface (see previous section).
 In the colloid fluctuation part, the fluctuations
of colloid heights and tilts are weighted by a Boltzmann factor which contains the energy
of the mean field solution (Euler--Lagrange equation) to the capillary problem  
with the boundary conditions set by the fluctuating contact line.
This decomposition is possible due to the fact that capillary wave Hamiltonian is Gaussian 
in the field $u$. In principle, it is possible to use this method also for the special
case of ellipsoidal colloids considered here. However, finding the mean field solution in such a 
geometry for arbitrary contact lines is rather cumbersome. To bypass this difficulty, 
we employed a trick adapted
from Ref.~\cite{rods} in which}
effective forces between rods on 
fluctuating membranes and films have been investigated. 
We extend the fluctuating interface height field $u(x,y)$ which enters the functional integral
for ${\cal Z}$ to the interior of the
ellipses $S_{i,\rm ref}$. Thus the measure of the functional integral for ${\cal Z}$ is  
extended by ${\cal D}u({\bf x})|_{{\bf x} \in S_{i,\rm ref}}$ and the
integration domain in the capillary wave Hamiltonian is enlarged to
encompass the whole $\mathbb{R}^2$. 
On the colloid surfaces, the interface height field is given by
the three phase contact line, $u(\partial S_{i,\rm ref}) \equiv f_i$.
We extend $u$ continuously to the interior of the circles $S_{i,\rm ref}$. 
{Such a continuation is not unique. However, the partition function remains
unchanged (up to a constant factor), if the energy cost of such a continuation 
is zero (as it is physically required since the interface is pinned to the ellipsoid
surface). This has to be insured by appropriate counterterms \cite {martin1, martin2}.
We choose the continuation:}
\be\label{artcoll}
u(S_{i,\rm ref})
\equiv
f_{i,\rm ext}(\xi_i,\eta_i)
= \sum_{m} \left( {P_{im}\, \frac{\cos(h m\xi_i)}{\cosh (m\xi_0)}\cos (m\eta_i) + 
Q_{im}\, \frac{\sinh (m\xi_i)}{\sinh (m\xi_0)}\sin (m\eta_i)} \right)
 \;,
\ee
where $\xi_i$ and $\eta_i$ are the elliptic coordinates with
respect to ellipse $S_{i,\rm ref}$.
The specific choice above is
convenient for the further calculations since $\nabla^2 f_{i,\rm ext}=0$
in $S_{i,\rm ref}\backslash \partial S_{i,\rm ref} $.
Extending the integration domain of the capillary wave Hamiltonian in Eq.~(\ref{Htot2}),
$\Omega=\mathbb{R}^2 \setminus\cup_i S_{i,\rm ref} \to \mathbb{R}^2$
generates an additional energy contribution $ -\mathcal{H}_{i,{\rm corr}}$
which has to be subtracted from the extended capillary wave Hamiltonian
$\mc{H}_{\rm cw}[\Omega \equiv \mathbb{R}^2]$. Therefore, the total Hamiltonian
reads:
\bea\label{Htotkard}
\mc{H}_{\rm tot}
&=&
\mc{H}_{\rm cw}+\sum_{i=1}^2[\mc{H}_{i,\rm b}+
\mc{H}_{i,\rm corr}]\;.
\eea
The correction Hamiltonian is calculated in App.~\ref{app:Hb}, and we recall the boundary 
Hamiltonian:
\bea\label{Hcorr1}
 -\mathcal{H}_{i,{\rm corr}}& =& \frac{\gamma \pi}{2} \sum_m m 
\left( P_{im}^2 \tanh (m\xi_0)  + Q_{im}^2 \coth (m\xi_0) \right)\;. \\
\mathcal{H}_{i,{\rm b}} &=&\frac{\gamma \pi}{2}(P_{i1}^2\tanh(\xi_0)+Q_{i1}^2\coth(\xi_0))\;.
\nonumber
\eea
In Eq.~(\ref{Hcorr1}) we have already omitted the contributions from the
gravitational term in $\mc{H}_{\rm cw}$ which are of order $(a/\lambda)^2 \ll 1$.

As in the previous section the 
partition function is written as a functional integral
over all possible configurations of the 
interface position $u$ and the boundary lines, expressed by $f_i$,
\begin{equation}\label{Zkard1}
\mathcal{Z}=\mathcal{Z}_0^{-1} \int \mathcal{D}u\,
\prod_{\rm i=1}^2
\int \mathcal{D}f_i
\prod_{{\bf x}_i \in  S_{i,\rm ref}}
\delta [u({\bf x}_{ i})-f_{i,\rm ext}({\bf x}_{ i})]
\exp \left\{-\frac{\mathcal{H}_{{\rm tot}}[f_i,u]}{k_{\rm B}T} \right\}
\;,
\end{equation}
where the product over the $\delta$-functions
enforces the pinning of the interface at the
positions of the colloids.
In contrast to Eq.~(\ref{Z1}), this product extends over all
${\bf x} \in S_{i,\rm ref}$ instead of $\partial S_{i,\rm ref}$, only.
The $\delta$-functions can again be expressed
by auxiliary fields $\psi_i$, now defined on the {\em two--dimensional} 
elliptical domains $S_{i,\rm ref}$
as opposed to the auxiliary fields of Sec.~\ref{sec:intflucpart} which are defined on the 
{\em one--dimensional} 
ellipses $\partial S_{i,\rm ref}$:
\begin{eqnarray}
\label{Zkard11}
 \mathcal{Z} &=&
\int \mc{D}u
\int \prod_{i=1}^2 \mc{D}\psi_i\int \mc{D}f_i\,
\exp\left\{
-\frac{\mathcal{H}_{{\rm tot}}[f_i,u]}{k_{\rm B}T}
+{\rm i}\int_{S_{i,\rm ref}}d^2x\,
\psi_i({\bf x})[u({\bf x})-f_{i,\rm ext}({\bf x})]
\right\}\;.
\end{eqnarray} 
%
Similarly to the evaluation of the fluctuation part, Sec.~\ref{sec:intflucpart}, we introduce
multipole moments $\Psi_{im}$ of the auxiliary fields by inserting
unity into ${\cal Z}$, Eq.~(\ref{Zkard11}):
\begin{eqnarray}
\label{unity}
\mathbb{1}
=\int \prod_{i=1}^2
\prod_{m } d{\Psi}_{im}^c d{\Psi}_{im}^s
\,\delta
\left(\Psi_{im}^c -\int_{S_{i,{\rm ref}}}d^2x\,(\cosh (m\xi)/\cosh (m\xi_0))\cos (m\eta)\, \psi_i({\bf x})\right)
 \nonumber\\
\times \,\delta
 \left(\Psi_{im}^s -\int_{S_{i,{\rm ref}}}d^2x\,(\sinh (m\xi)/\sinh (m\xi_0))\sin (m\eta)\, \psi_i({\bf x})\right)
\;.
\end{eqnarray}
In contrast to the evaluation of the fluctuation term in Sec.~\ref{sec:intflucpart}, there
will be constraints on the lowest multipoles which contribute to ${\cal Z}$. 
To see this
we note that the Hamiltonian 
$\mathcal{H}_{{\rm tot}}$ does not depend on the boundary monopole moments
$P_{i0}$ and the dipole moments $P_{i1}$ {(through a cancellation between
${\cal H}_{i,{\rm b}}$ and ${\cal H}_{i,{\rm corr}}$)}, and the only dependence of ${\cal Z}$ on these
moments is through the constraint function $f_{i,\rm ext}$. Recalling the definition of the
integration measure ${\cal D}f_i$ for the {two  boundary conditions (B) and (C) and}
 performing the integration over $P_{i0}$ (B) and $P_{i0}$ and $P_{i1}$ (C),
we immediately find
\begin{equation}
\label{constraints}
{\cal Z} \sim \left\{ 
\begin{array}{ll} 
{\displaystyle \int \prod_{i=1}^2 \prod_{m } d{\Psi}_{im}^s \, d{\Psi}_{im}^c \dots \delta  
(\Psi_{i0}^c ) \dots} & \mbox{case (B)}
 \\
{\displaystyle \int \prod_{i=1}^2 \prod_m d{\Psi}_{im}^s \,d{\Psi}_{im}^c 
\dots \delta ( \Psi_{i0}^c)  \,
\delta ( \Psi_{i1}^c)\, \delta ( \Psi_{i1}^s) \dots \qquad }& \mbox{cases (C)}
\end{array}
\right.
\end{equation}
Having noticed these constraints on the auxiliary fields, we proceed by
integrating 
over the 
field $u$ in Eq.~(\ref{Zkard11}):
\begin{eqnarray}
\label{Zkard2}
 \mathcal{Z} &=&
\int \prod_{i=1}^2 \mc{D}\psi_i\int \mc{D}f_i\,
\exp\left\{
-\frac{k_{\rm B}T}{2\gamma}\sum_{i,j=1}^2
\int_{ S_{i,\rm ref}}d^2x_i \int_{ S_{j,\rm ref}}d^2x_j\,
\psi_i ({\bf x}_i)\,G(|{\bf x}_i-{\bf x}_j|)\,\psi_j({\bf x}_j)
\right. \nonumber\\ &&\left.
-\frac{1}{k_{\rm B}T} \left(\mc{H}_{i,\rm{b}} + \mc{H}_{i,\rm{\rm{corr}}}\right)
- {\rm i}\sum_{i=1}^2\int_{S_{i,\rm ref}}{\rm d}^2x\,\psi_i({\bf x})f_{i,\rm ext}({\bf x})
\right\}\;,
\end{eqnarray} 
where -- as in Eq.~(\ref{Zaux}) -- 
$G$ is the Greens function of the capillary wave Hamiltonian.
A somewhat longer calculation shows that ${\cal Z}$ can be split
into into an interaction part (coupling the auxiliary multipole moments
$\Psi^{c[s]}_{im}$, $P_{im}$ and $Q_{im}$ for different colloid labels $i$), a self--energy part
(depending on $\Psi^{c[s]}_{im}$, $P_{im}$ and $Q_{im}$ 
for each value of $i$ separately) and a remainder (the sum
of boundary and correction Hamiltonian):
\begin{eqnarray}
 \mathcal{Z} &=&  \int \prod_{i=1}^2 \prod_m d\Psi_{im}^c d\Psi_{im}^s \int \mc{D}f_i\,
  \exp\left\{
-\frac{k_{\rm B}T}{2\gamma}\left(\mc{H}_{\rm int}[\Psi^{c[s]}_{1m},\Psi^{c[s]}_{2m}] +
  \mc{H}_{i,\rm self}[\Psi^{c[s]}_{im}]\right) \right\} \times \nonumber \\
 \label{Zkard3} 
  &&\qquad \qquad 
 \exp\left( \frac{1}{k_{\rm B}T}
 \left(\mc{H}_{i,\rm{b}} + \mc{H}_{i,\rm{\rm{corr}}}\right) -{\rm i} \sum_m(\Psi_{im}^c P_{im} + \Psi_{im}^s Q_{im}) 
\right)
\end{eqnarray}
The interaction part
\begin{eqnarray}
\label{kardint} 
\mc{H}_{\rm int}  &=&  2\int_{S_{1,\rm{ref}}} d^2x_1\,\int_{S_{2,\rm{ref}}} d^2x_2 
\psi_1 ({\bf x}_1)G_{\rm int}(|{\bf x}_1 - {\bf x}_2|)\psi_2({\bf x}_2)
\end{eqnarray}
{turns out to be a bilinear form in the auxiliary multipole moments; this is shown 
using the already used multipole expansion of the Greens function
 $G(|{\bf x}_1-{\bf x}_2|) 
\simeq -\ln(\gamma_{\rm e}|{\bf x}_1-{\bf x}_2|/2\lambda_c)$ (valid for $d \gg a$)
which is
presented in App.~\ref{app:schwing} in more detail}. {This bilinear form reads
\bea
\label{kardint1} 
 2\pi\;\mc{H}_{\rm int}  &=&  
     \left(
      \begin{array}{c}
        {\bf\widehat{\Psi}}_1^c\\
        {\bf\widehat{\Psi}}_1^s
      \end{array}
    \right)
    ^{\rm T}
    \left(\begin{array}{cc}
        \wh{ \bf G}_{\rm int}^{cc} & \wh{\bf{G}}_{\rm int}^{sc} \\
        \wh{\bf{G}}_{\rm int}^{sc} & \wh{\bf{G} }_{\rm int}^{ss}
      \end{array}\right)
    \left(
      \begin{array}{c}
        {\bf \widehat{\Psi}}_2^c\\
        {\bf \widehat{\Psi}}_2^s
 \end{array}
\right)\;,
\eea
where the submatrices $\wh{ \bf G}_{\rm int}^{cc}$, $\wh{ \bf G}_{\rm int}^{sc}$ and $\wh{ \bf G}_{\rm int}^{ss}$
have already been encountered in the calculation of the fluctuation part and are given by 
Eqs.~(\ref{int-cc00})--(\ref{sc}).
}
%
 The self--energy part (different from the corresponding one in the calculation of the fluctuation part) 
is evaluated in App.~\ref{app:se}, with the result
\begin{eqnarray}
\label{sekard}
 \mathcal{H}_{i,{\rm self}} &=& - \frac{\ln\left(\gamma_{\rm e} (a+b)/8\lambda_c\right)}{2\pi}{\Psi_{i0}^c}^2 \nonumber \\
&&+ \frac{1}{\pi}\sum_{m>0} \frac{1}{m}\left(\frac{{\Psi_{im}^c}^2}{1+\tanh (m\xi_0)}+\frac{{\Psi_{im}^s}^2}{1+\coth (m\xi_0)}\right)
\end{eqnarray}
Combining Eqs.~(\ref{Zkard3}), (\ref{kardint}), and (\ref{sekard}), the partition
function can be written as
\begin{eqnarray}\label{Zkard6}
  \mc{Z}
  &=&
 \int \prod_{i=1}^2\prod_m
    \mc{D}\Psi_{im} \mc{D}f_i\,
  \exp\left\{-
    \frac{k_{\rm B}T}{2\gamma}
    \left(
      \begin{array}{c}
        {\bf\wh{{\Psi}}}_1\\
        {\bf\wh{{\Psi}}}_2
      \end{array}
    \right)
    ^\dagger
    \left(\begin{array}{cc}
        \wh{ \bf H}_{\rm self} & \wh{\bf{ H}}_{\rm int} \\
        \wh{\bf{H}}_{\rm int} & \wh{\bf{H} }_{\rm self}
      \end{array}\right)
    \left(
      \begin{array}{c}
        {\bf\wh{{\Psi}}}_1\\
        {\bf\wh{{\Psi}}}_2
 \end{array}
\right)
\right\}\;,
\end{eqnarray}
where the vectors 
${\bf{\Psi}}_i=
(\Psi_{i0}^c,  P_{i0},
\Psi_{i1}^c,P_{i1}, 
\Psi_{i1}^s,Q_{i1}
,\dots)$ -- in contrast to $\wh{\bf{\Psi}}_i$
in Sec.~\ref{sec:intflucpart} --
contain all involved auxiliary
and boundary multipole moments.
The elements of
 the matrix ${\bf{H}}$ describe the coupling of
these multipole moments, where the self-energy block
couples multipoles defined on the same ellipses $S_{i,\rm ref}$. 
{Thus the diagonal part of the self energy matrix $\wh{ \bf H}_{\rm self}$ 
can be read off Eq.~(\ref{Hcorr1}) and Eq.~(\ref{sekard}) while the off--diagonal part is determined
by the term $-{\rm i} \sum_m(\Psi_{im}^c P_{im} + \Psi_{im}^s Q_{im})$ in Eq.~(\ref{Zkard3}).}
The elements of the interaction matrix $\wh{ \bf H}_{\rm int}$
are determined by the interaction energy $\mc{H}_{\rm int}$ in
Eqs.~(\ref{kardint}) and (\ref{kardint1}) 
and couple the auxiliary multipole moments of
different colloids. All matrix elements
representing couplings of other multipoles are zero.
Similar as in Eq.~(\ref{Zauxmom}),
the exponent in Eq.~(\ref{Zkard6}) is a bilinear
form, however, here combined for all
types, boundary multipole moments $P_{im}$, $Q_{im}$  and 
 auxiliary multipoles $\Psi_{im}^c$, $\Psi_{im}^s$.
The computation of the partition function
amounts to the calculation of  $\det\wh{\bf H}$.
{Again this is found as a series expansion in $a/d$, and we may define
a similar expansion for the free energy $\mc{F}_{\rm in+coll}=-(k_{\rm B}T)\ln \det\wh{\bf H}/2$
as before in Eq.~(\ref{Fflucexp}): 
\bea\label{Fflucexp1}
\mc{F}_{\rm in+coll}(d)
&=&
k_{\rm B}T
\sum_{n}f_{2n}^{\rm in+coll}
\,\left(\frac{1}{d}\right)^{2n}\;.
\eea
The leading coefficients in case (B) (inclusion of fluctuations in the ellipsoids' vertical positions) 
are given by:
\bea
 \label{bres}
   f_0^{\rm in+coll}=f_2^{\rm in+coll} &=& 0 \;, \\
   f_4^{\rm in+coll} &=&{\displaystyle -\frac{1}{2^8} 
 \left[\,(a^2-b^2)^2\, \cos(2\theta_1 + 2\theta_2) + (a+b)^4  \,\right]}\;. \nonumber
\eea
In case (C) (inclusion of fluctuations in the ellipsoids' vertical positions and tilt angles with respect
to the interface) the leading coefficients are:
\bea
 \label{cres}
   f_{2n}^{\rm in+coll} &=& 0 \qquad (n=0 \dots 3) \;, \\
   f_8^{\rm in+coll} &=&-\frac{9}{2^{16}}
\left[\, (a^2-b^2)^4\,\cos(4\theta_1 + 4\theta_2) + (a+b)^8 \,\right]\;. \nonumber
\eea
}
In contrast to the calculation before, the different leading power laws 
for the different cases (B) and (C)
can be understood easily. 
We note that the interaction between the auxiliary multipoles $\Psi_{1m}^{(c,s)}$
 and $\Psi_{2n}^{(c,s)}$ 
in $\mc{H}_{\rm int}$, Eq.~(\ref{kardint}),
scales like $(a^{\prime}/4d)^{m+n}$.
 {After calculating the determinant}, the leading order of the total fluctuation induced
force between the two colloids is determined by the
first non--vanishing auxiliary multipole moment $\Psi_{im'}^{(c,s)}$ and (as follows
from $\det\wh{ \bf H}$) gives rise to a term in the free energy $ \propto 1/d^{4m'}$ (for $m'>0$)
or $ \propto \ln \ln d$ (for $m'=0$). 
As explained in the beginning of this subsection, the different
boundary conditions lead to certain constraints on the auxiliary multipoles:
According to Eq.~(\ref{constraints}), the leading term in $F(d)$
arises from a monopole-monopole interaction of the auxiliary field in case (A), 
from a dipole-dipole interaction in case (B),
and from a quadrupole-quadrupole interaction in case (C).
The constraints of vanishing auxiliary monopole and dipole moments (as in (C)) result from the 
independence of $\mc{H}_{\rm tot}$
of the boundary  monopole and dipole moments and this is only captured correctly 
by the inclusion of
the correction Hamiltonian $\mc{H}_{\rm corr}$. 
{The Casimir attraction is maximal if the major axes of both ellipses are oriented parallel,
regardless of the orientation of the distance vector joining their centers. This is 
a peculiarity in two dimensions, as can be seen also by the general multipole expansion
of the interaction between two arbitrary charge distributions in two--dimensional electrostatics.  }
\subsection{Limiting cases}

In the limit $a=b$ (colloids with circular contact line such as disks and spheres) our results for
the cases (A)--(C) reduce to the results reported in Refs.~\cite{martin1,martin2}. In the limit
$b\to 0$ (colloidal rods or needles with vanishing thickness) we can compare our result for case (C)
(fluctuating colloid heights and tilts) to Ref.~\cite{rods}. There it has been found that the effective free 
energy asymptotically varies $\propto d^{-4}$ with a coefficient given by Eq.~(\ref{bres}), i.e.
by the result of case (B) (colloid height fluctuations only). The derivation in Ref.~\cite{rods}
suggests that the perturbative treatment employed in our approach should be amended by corrections
in the integration measure over the tilts. In our cases, this measure is simply given by 
$dP_{i1}dQ_{i1}$, the product of the measures for the cosine and sine dipole moments of the contact line.
{In order to check the validity of this approximation for the measure, we recalculated 
the partition function of Ref.~\cite{rods} by perfroming the integrations over the
auxiliary dipole moments and their conjugate variables, which results in a final integral 
over the tilts (with the general measure) weighted by an exponential function in the tilts.
If $\gamma a^2/(k_{\rm B}T) \gg 1$, the denominator of the measure in this integral can be 
expanded in terms of the tilt angles since the exponential function decays much faster
than the measure and thus determines the convergence of the integral. In this way, one sees that
the higher--order terms in the dipole tilt measure do not provide another leading
behaviour in $1/d$ in the partition 
function compared to the leading quadrupole--quadrupole interaction which arises in our 
perturbative picture.  
A breakdown of our perturbative treatment can be expected
if the length of the rod $a$ approaches the molecular length scale. This coincides with a simultaneous
breakdown of the simple capillary wave picture underlying our analysis.  
}
\section{Summary} 
\label{summary}
The restrictions that two colloids
trapped at a fluid interface impose
on the thermally excited interfacial fluctuations (capillary waves)
by their sheer presence lead to a thermal Casimir interaction.
{We have obtained an explicit account for the effect of colloidal anisotropy on the form
of the Casimir interaction by studying ellipsoidal (spheroidal) colloids with arbitrary 
aspect ratio. For the case of fixed colloids and fixed contact lines, the problem is equivalent
to the ``standard" Casimir problem for a scalar, Gaussian field in two dimensions with 
Dirichlet boundary conditions on the  colloid surface. In an expansion in $1/d$, the inverse
center--to--center distance between the colloids, the leading term in the Casimir interaction energy is
found to be attractive, isotropic in the interface plane and slowly varying $\propto \ln\ln d$
(see Eq.~(\ref{ares})).
Anisotropies appear in higher orders in $1/d$ and become important when the closest surface--to--surface
distance between the colloids becomes small (see Fig.~\ref{figures}).   
}

{
If fluctuations in the colloids' vertical position are permitted, the asymptotics of the
Casimir interaction energy changes to a behaviour $\propto d^{-4}$ (see Eq.~(\ref{bres})). 
In this case, anisotropies are present in the leading term but the interaction remains attractive
for all orientations. If furthermore fluctuations of the colloids' orientation with respect to the interface
normal are allowed, the asymptotics changes to a behaviour $\propto d^{-8}$ (see Eq.~(\ref{cres})).
Interestingly, this change of leading order in the asymptotics of the Casimir energy depending on the
type of permitted colloid fluctuations holds for arbitrary aspect ratios. This leads to the 
speculation that this might be a general feature holding for arbitrary colloid shape.  
}

{
In our approach, the Casimir interaction can be understood as the interaction between fluctuating multipole
moments of an auxiliary charge density--like field defined on the area enclosed by the contact lines.
These fluctuations are coupled to fluctuations of multipole moments of the contact line position which are
a due to the possibly fluctuating colloid height and tilts. Therefore, the system can be viewed as
an example for the Casimir effect with fluctuating boundary conditions. Such fluctuating boundary conditions
appear to be difficult to be realizable in three--dimensional systems such as the standard system of
charged metallic objects subjected to vacuum fluctuations of the electromagnetic field. 
}

{Experimentally, the detection of the Casimir interaction at a fluid interface 
appears to be possible if competing interactions, especially van--der--Waals and static
capillary interactions, are sufficiently weakened. Van--der--Waals interactions 
are also strongly attractive at small distances, but can be modified by an appropriate
core--shell structure of the colloids or by using flat, disk--like particles
(we refer to a longer discussion of this issue in Ref.~\cite{martin1}). 
Capillary interactions are very strong for ellipsoidal colloids of micrometer size and 
 with contact angle different from $\pi/2$ since the equilibrium contact line in this case
is already undulated and gives rise to static deformations of the surrounding interface
 \cite{meandmartin}. These capillary interactions can be minimized by either using
truly nanoscopic ellipsoids or synthesizing Janus particles with a contact line which is flat on
a nm level. Despite the great advances in particle synthesis over the last years, this 
appears to be still a big challenge.} 
 
{\bf Acknowledgment:} 
The authors thank the German Science Foundation for financial
support through the Collaborative Research Centre (SFB-TR6) ``Colloids in
External Fields", {project N01}.

\begin{small}
\appendix
\section{Confocal Elliptic Coordinate System} \label{elliptic_coordinates}
Confocal elliptic coordinates {$(\xi,\eta)$} 
are planar orthogonal coordinates formed by confocal ellipses or hyperbolae. 
The foci are located on the $x$-axis of the Cartesian coordinates, separated by {$a'$}. 
The relation to Cartesian coordinates 
is defined by
\begin{eqnarray}
x = \frac{a^{\prime}}{2}\cosh(\xi)\cos(\eta) \;, \nonumber\\
y = \frac{a^{\prime}}{2}\sinh(\xi)\sin(\eta) \;,
\end{eqnarray}
and  the scale factors are found as
\begin{equation}
h_{\xi} = h_{\eta} =\sqrt{ \frac{{a^{\prime}}^2}{2}\left(\cosh(2\xi)-\cos(2\eta)\right)} \;.\nonumber\\
\end{equation}
$\xi$ and $\eta$ are called elliptic radius and elliptic angle, respectively. 
In this coordinate system, $\xi = \xi_0$ represents the equation of an ellipse with axes $a,\,b$ ($a>b$). 
The elliptic radius and the distance $a^{\prime}$ between the foci are given in terms of the
ellipse principal axes by
\begin{eqnarray}
\xi_0 = \frac{1}{2}\ln\left(\frac{a+b}{a-b}\right) \;, \nonumber\\
a^{\prime} = (a^2-b^2)^{\frac{1}{2}}\;.
\end{eqnarray}
Therefore, for a circle we have $\xi_0\rightarrow \infty$ and for a line, $\xi_0\rightarrow 0$.
\section{The Boundary and Correction Hamiltonians $\mc{H}_{\rm i,b}$ and $\mc{H}_{\rm i,\rm corr}$ }
\label{app:Hb}
\begin{itemize}
\item[{\bf i)}]
The boundary Hamiltonain for the case of a pinned contact line (Janus ellipsoids) is governed 
by the difference in projected meniscus area, $\Delta A_{\rm proj}$ (Eq.~(\ref{aproj}). 
If the ellipsoid is tilted in the $xz$-plane by a small angle $\alpha_i$, this area is given by
\begin{equation}
\label{project} 
\Delta A_{\rm proj}^{xz} \approx \frac{\pi}{16} \alpha_i^2 {a^{'}}^2 \sinh (2\xi_0) \;.
\end{equation}
{The contact line position $u|_{\partial S_{i,{\rm ref}}}$ is easily determined by using
a coordinate system rotation from the $xz$-plane to  the $x^{'}z^{'}$-plane, where the $x^{'}$-- 
and the $z^{'}$--axis coincide with the major and minor axis of the tilted ellipsoid. The contact line
is located at $z^{'}=0$ in the new coordinate system. Since $z^{'} \approx z - \alpha_i x$, we find
that $u|_{\partial S_{i,{\rm ref}}} = z|_{z'=0} \approx  \alpha_i x$ and therefore
$u|_{\partial S_{i,{\rm ref}}} \approx\alpha_i (a^{'}/2) \cosh (\xi_0) \cos (\eta_i)$. 
Thus, in the multipole expansion of the tilted contact line (Eq.~(\ref{conlin})), only a dipole term appears with
the dipole moment given by:} 
\begin{equation}
\label{alpha}
P_{i1}=\alpha_i \frac{a^{'}}{2} \cosh (\xi_0)\;.
\end{equation}
Inserting eq.~(\ref{alpha}) into (\ref{project}) we obtain
\begin{equation}
 \Delta A_{\rm proj}^{xz} = \frac{\pi}{2} P_{i1}^2\tanh (\xi_0)  \;.
\end{equation}
Applying the same arguments for tilts in the $yz$-plane, we can express the boundary Hamiltonian in the 
small--tilt approximation by
\begin{eqnarray}
 \mathcal{H}_{i,{\rm b}}&=& \gamma \Delta A_{\rm proj} \nonumber \\
&=& \frac{\gamma\pi}{2} (P_{i1}^2\tanh (\xi_0) + Q_{i1}^2\coth (\xi_0))\;.
\end{eqnarray}
\item[{\bf ii)}]
{The correction Hamiltonian, which is introduced in Eq.~(\ref{Htotkard}), is determined by minus
the surface energy of the meniscus piece $u|_{S_{i,{\rm ref}}} \equiv f$ (Eq.~(\ref{artcoll})) extended into the ellipses enclosed
by the reference contact lines:} 
\begin{eqnarray}
 -\mathcal{H}_{i,{\rm corr}} &=& \frac{\gamma}{2} \int_{S_{i,{\rm ref}}} d^2x 
\left[(\nabla u)^2+\frac{u^2}{\lambda_c^2}\right] \nonumber\\
& \stackrel{\lambda_c \rightarrow \infty}{=} & \frac{\gamma}{2} \int d^2x (\nabla f)^2\;.
\end{eqnarray}
Thus,
\begin{eqnarray}
 \label{energy_si}
 -\mathcal{H}_{i,{\rm corr}} & = & \frac{\gamma}{2} \int_0^{\xi_0}\int_0^{2\pi} d\xi_i d\eta_i \left((\partial_{\xi_i}f)^2+(\partial_{\eta_i}f)^2\right) \nonumber\\
&=& \frac{\gamma \pi}{2} \sum_m m \left(  P_{im}^2 \tanh (m\xi_0) +  Q_{im}^2 \coth (m\xi_0)\right)\;.
\end{eqnarray}
\end{itemize}
\section{Expansion of Green's function in elliptic coordinates}\label{app:schwing}
In this appendix we derive the multipole expansion of the Green's function
 $G(|\mathbf x |)\approx-(1/2\pi)\ln (\gamma_{\rm e} |\mathbf x | / 2\lambda_{c})$ 
between two charged elliptic regions (charges are generated by auxilliary fields $\psi_{i}$).
 This Green's function  gives the correlation between two points residing either 
on the same ellipse or different ellipses. \\
1) In the case that  ${\mathbf x}_1$ and ${\mathbf x}_2$ are located on the same ellipse,
 the Green's function expansion has been given in Ref.~\cite{feshbach}:
\begin{equation}
 \label{gself}
2\pi G (|{\mathbf x_1}-{\mathbf x_2}|) = -\ln\left({\frac{\gamma_{\rm e} a^{'} e^{\xi_0}}{8\lambda_c}}\right)
+ 2\sum_{n=1}^\infty
\frac{e^{-n \xi_0}}{n} \left[ \cosh (n\xi_0) \cos (n \eta_1) 
\cos (n \eta_2) + \sinh (n\xi_0) \sin (n\eta_1) \sin (n\eta_2) \right] 
\end{equation}
2) {The case that the two points are located  on different ellipses, i.e. 
$\mathbf x_1= \mathbf r_1$ and $\mathbf x_2 = \mathbf d+ \mathbf r_2$, which 
furthermore possess an arbitrary orientation in the plane (expressed by the angles
$\theta_1$ and $\theta_2$, see Fig.~\ref{plane}) is more difficult.
We start with a general Taylor expansion of the Green's function:}
\begin{equation}
\label{green-expansion}
-\frac{1}{2\pi}\ln \left(\frac{\gamma_{\rm e}|{\mathbf d+\mathbf r_2-
 \mathbf r_1}|}{2\lambda_c}\right) = -\frac{1}{2\pi}
\ln\left(\frac{\gamma_{\rm e} d}{2\lambda_c}\right)
 -\frac{1}{2\pi}\left.\sum\limits_{ j_1,j_2=0 \atop j_1+j_2 \geq1}
\frac{\left(-\mathbf r_1.\mathbf\nabla\right)^{j_1}}{j_1!}
\frac{\left(\mathbf r_2. \mathbf \nabla\right)^{j_2}}{j_2!}\ln r \right|_{r=d}\;.
\end{equation}
{On the other hand, we can perform this expansion using complex variables $z=x+{\rm i}y$.
The expansion of the logarithm is given by:}
\begin{equation}
\label{complex-taylor}
 \ln(z-z^{'})=\sum_{j=0}^\infty \frac{1}{j!} (-z^{'}\partial_z)^j \ln z \;.
\end{equation}
The real part of eq.~(\ref{complex-taylor})
 is the expansion of the real logarithm, of course:
\begin{equation}
\label{realpart}
 \ln|\mathbf{r-r^{'}}|=\mathrm{Re}
\sum_{j=0}^\infty -\frac{1}{j}\frac{(z^{'}z^{*})^j}{|z|^{2j}} \;.
\end{equation}
By comparing Eq.~(\ref{realpart}) and the Taylor expansion of
 $\ln |\mathbf r-\mathbf r^{'}|$ in real space as in Eq.~(\ref{green-expansion}), we find
\begin{equation}
\label{real-imag} 
\frac{(-\mathbf r^{'}.\mathbf\nabla)^{j}}{j!}\ln r =
 -\frac{1}{j} {\mathrm {Re}}\frac{(z^{'}z^{*})^j}{|z|^{2j}} \;.
\end{equation}
Introducing complex derivative operators $\zeta_-=\partial_z$ , 
$\zeta_+=\partial_{z^*}$ and using $\zeta_+ \zeta_- \ln r = \zeta_- 
\zeta_+ \ln r = 0 $, we have
\begin{equation}
\label{zeta-effect}
\zeta_\pm^j \ln r=\zeta_\pm^j \ln |z|= \frac{1}{2}(-1)^{j-1} (j-1)!\;
\left\{
\begin{array}{l}
z^j/|z|^{2j}\\z^{*j}/|z|^{2j}
\end{array}
\right. \;.
\end{equation}
Identifying eq.~(\ref{real-imag}) and ~(\ref{zeta-effect}) we obtain
\begin{equation}
\label{real-imag-zeta}
\frac{(\mathbf r^{'}.\mathbf\nabla)^{j}}{j!}\ln r 
= \frac{1}{j!}\left((z^{'*})^j\zeta_+^j + z^{'j} \zeta_-^j\right)\ln r \;.
\end{equation}
By inserting eq.~(\ref{real-imag-zeta}) into eq.~(\ref{green-expansion}),
 we find the Green's function expansion in terms of complex variables $z_i$
\begin{equation}
\label{green-expansion-complex}
G |{\mathbf x_2}-{\mathbf x_1}| = -\frac{1}{2\pi}\ln\left(\frac{\gamma_{\rm e} d}
{2\lambda_c}\right)+\frac{1}{2\pi}\sum_{\tiny{\begin{array}{c}
 j_1,j_2=0\\ j_1+j_2\geq1
\end{array}}} \frac{(-1)^{j_2}}{j_1+j_2} 
\left(\begin{array}{c}
  j_1+j_2\\ j_1 
\end{array}\right)
\frac{1}{d^{j_1+j_2}} {\mathrm {Re}}[z_1^{j_1}z_2^{j_2}] \;.
\end{equation}
This general expansion can be used in any coordinate system.
{In the special case of elliptic coordinates, $z=(a^{'}/2)e^{-i\theta}\cosh(\xi+i\eta)$,
 where $\theta$ is the in-plane rotation angle of the ellipse major axis with respect to a fixed
$x$--axis. (For the configuration of arbitrarily oriented ellipses, the $x$--axis is given by the
line joining their centers, see Fig.~\ref{plane}).}  In order to express 
Eq.~(\ref{green-expansion-complex}) in elliptic coordinates,
 one needs $z^{j}$ which can be found by applying the binomial expansion:
\begin{equation}
\label{binomial}
z^j= \left( \frac {a^{'} e^{-i\theta}}{4} \right)^j 
\sum_{k=0}^{j} {j \choose k} e^{(2k-j)(\xi+i\eta)} \;.
\end{equation}
Using the above expansion, 
Eq.~(\ref{green-expansion-complex}) becomes:
\begin{eqnarray}
\label{elliptic-expansion}
G(|\mathbf x_1 - \mathbf x_2|) &=& -\frac{1}{2\pi}
\ln\left(\frac{\gamma_{\rm e} d}{2 \lambda_c}\right) \nonumber \\
&& +\frac{1}{2\pi} \sum_{j_1,j_2=0 \atop j_1+j_2 \geq 1}
\sum_{k_1=0}^{j_1}\sum_{k_2=0}^{j_2} 
\frac{(-1)^{j_2}}{j_1+j_2}{j_1+j_2 \choose j_1}{j_1 \choose k_1}{j_2 \choose k_2}
{\left(\frac{a^{'}}{4d}\right)}^{j_1+j_2} \nonumber \\
&&\times \exp(\xi_1(2k_1-j_1)+\xi_2 (2k_2-j_2)) \nonumber \\ 
&& \times \cos(j_1\theta_1+j_2\theta_2+\eta_1(j_1-2k_1)+\eta_2(j_2-2k_2))\;.
\end{eqnarray} 
{The aim is to rewrite this fourfold sum over $j_1,j_2,k_1,k_2$ as an expansion into
multipole coefficients $\cos(m\eta_1)\cos(n\eta_2)$ and $\sin(m\eta_1)\sin(n\eta_2)$
with $m,n \ge 0$. To that end, we define $m=|j_1-2k_1|$ and $n=|j_2-2k_2|$.
The possibility that $j_i-2k_i$ may be positive as well as negative makes
it necessary to consider the following cases:}
{\bf (i)} ``$m=0,\;n=0$'', {\bf (ii)}``$m=0,\; n\neq 0$'' 
or ``$m \neq 0,\; n=0$'' and {\bf (iii)} ``$m \neq 0,\; n \neq 0$''. \\
\begin{itemize}
\item[{\bf (i)}] {``$m=0,\;n=0$'': Two auxiliary variables $l_1,l_2$ are introduced through 
$j_i=2l_i, k_i = l_i$ ($i=1,2$). 
The such constrained sum in Eq.~(\ref{elliptic-expansion}) reduces to }
$$
\sum_{l_1,l_2=0 \atop l_1+l_2 \geq 0} \frac{1}{2(l_1+l_2)}
{2(l_1+l_2) \choose 2l_1}{2l_1 \choose l_1}{2l_2 \choose l_2}
{\left(\frac{a^{'}}{4d}\right)}^{2(l_1+l_2)}\cos(2l_1\theta_1+2l_2\theta_2) \;.
$$
Relabelling $l=l_1+l_2$ and $l^{\prime}=l_1$, the above sum is rewritten as
\begin{equation}
\label{m0n0} 
\sum_{l=1}\sum_{l^{\prime}=0}^l \frac{1}{2l}
{2l \choose 2l^{\prime}}{2l^{\prime} \choose l^{\prime}}{2(l-l^{\prime}) \choose l-l^{\prime}}
{\left(\frac{a^{'}}{4d}\right)}^{2l}\cos(2l^{\prime}\theta_1+2(l-l^{\prime})\theta_2) \;.
\end{equation}
\\
\item[{\bf (ii)}] {``$m=0,\; n > 0$'': Here, $l_1$ is introduced as above through
$j_1=2l_1$ and $k_1=l_1$. We distinguish the two  cases $j_2-2k_2>0$ and
$j_2-2k_2<0$ via the choice of $l_2$ through
$j_2=n+2l_2$ and $k_2=l_2$ vs. $k_2=n+l_2$. 
Adding up these two cases in the constrained sum (\ref{elliptic-expansion}), 
and performing a relabelling analogous to 
the one leading to expression
(\ref{m0n0}) ($l=l_1+l_2$, $l^{\prime}=l_1$)  we obtain:}
\begin{eqnarray}
 2\sum_{l=0}\sum_{l^{\prime}=0}^l (-1)^n
\frac{\Gamma(n+2l)}{{l^{\prime}}
^2 (l-l^{\prime})!\left(n+
l-l^{\prime}\right)!} \left(\frac{a^{'}}{4d}\right)^{n+2l}
 &&\{ \cos(2l^{\prime}\theta_1 +(n+2(l-l^{\prime}))\theta_2)
\cosh (n\xi_2) \cos (n\eta_2) \nonumber\\
& + & \sin (2l^{\prime}\theta_1 +(n+2(l-l^{\prime}))\theta_2)
 \sinh (n \xi_2) \sin (n \eta_2) \} \;. \nonumber \\
\end{eqnarray}
Similarly we obtain for ``$m > 0, \; n=0$ '':
\begin{eqnarray}
  2\sum_{l=0}\sum_{l^{\prime}=0}^{l}\frac{\Gamma(m+2l)}{l^{\prime}
!{(l-l^{\prime})!}^2\left(m+l^{\prime}\right)!}
 \left(\frac{a^{'}}{4d}\right)^{m+2l} 
 && \{ \cos((m+2l^{\prime})\theta_1 + 2(l-l^{\prime})\theta_2)\cosh (m\xi_1) \cos (m\eta_1) \nonumber \\
& + &\sin ((m+2l^{\prime})\theta_1 + 2(l-l^{\prime})\theta_2) \sinh (m \xi_1) \sin (m \eta_1) \} \;. \nonumber \\
\end{eqnarray}
\item[{\bf (iii)}] {``$m> 0,\; n> 0$'': Similarly to the previous cases, $j_1$ and
$j_2$ are introduced through $j_1=m+2l_1$ and $j_2=n+2l_2$.
The four cases of possible sign combinations of  $j_i-2k_i$ ($i=1,2$) are
taken into account by the relation sets 
``$k_1=l_1 \;, k_2=l_2$'', ``
$k_1=l_1\;,k_2=n+l_2$'', ``$k_1=m+l_1\;,k_2=l_2$'',``$k_1=m+l_1,\;k_2=n+l_2$''.
 Adding up these four cases in the constrained sum (\ref{elliptic-expansion}) and taking 
advantage of the addition and subtraction
 relations between hyperbolic functions and then relabelling as before
 ($l=l_1+l_2$, $l^{\prime}=l_1$) we find:}
\begin{eqnarray}
 4\sum_{l=0}\sum_{l^{\prime}=0}^l (-1)^n\frac{\Gamma(m+n+2l)}
{\left(m+l^{\prime}\right)!l^{\prime}!
\left(n+l-l^{\prime}\right)!\left(l-l^{\prime}\right)!}
 \left(\frac{a^{'}}{4d}\right)^{m+n+2l}\nonumber\\
 ( \cos(\Theta)\cosh( m\xi_1) \cos (m\eta_1)
 \cosh (n\xi_2) \cos (n\eta_2) \nonumber\\
+ \sin (\Theta) \sinh (m \xi_1) \sin (m \eta_1) 
\sinh (n \xi_2) \sin (n \eta_2)) \;,
\end{eqnarray}
where $\Theta=(m+2l^{\prime})\theta_1+(n+2(l-l^{\prime}))\theta_2$.\\
\end{itemize}
{The cases considered above may be combined into a single expression such that the 2D 
Green's function in elliptic coordinates reads:}
\begin{eqnarray}
\label{final_exp} 
G(|\mathbf x_1-\mathbf x_2|) & = & -\frac{1}{2\pi}
\ln\left(\frac{\gamma_{\rm e} d}{2\lambda_c}\right) + \frac{1}{2\pi} \sum_{m,n,l=0}
 \left(\frac{a^{'}}{4d}\right)^{m+n+2l} \\
&& \times \{ A_{mnl}^c \left[ \cos (m\eta_1) \cosh (m\xi_1) 
\cos (n\eta_2) \cosh (n\xi_2) - \sin (m\eta_1) \sinh (m\xi_1) 
\sin (n\eta_2) \sinh (n\xi_2)\right] \nonumber \\
&& + A_{mnl}^s \left[ \sin (m\eta_1) \sinh (m\xi_1) \cos (n\eta_2) \cosh (n\xi_2) +
 \cos (m\eta_1)  \cosh (m\xi_1) \sin (n\eta_2) \sinh (n\xi_2)\right]\} \;. \nonumber
\end{eqnarray}
In Eq.~(\ref{final_exp}), the coefficients $A_{mnl}^{c \choose s}$ are given by
\begin{equation}
 \label{a_mnl}
 A_{mnl}^{c \choose s} = A_{mnl}^{c \choose s} (\theta_1,\theta_2)= 2^{H_n+H_m} \sum_{l^{\prime}=0}^l C_{ll^{\prime}}^{mn}  {\cos (\Theta) \choose \sin (\Theta)} \;,
\end{equation}
where $H_j$ is the discrete step function; $ H_j = 1-\delta_{j0}$, and
\begin{equation}
 \label{c_llmn}
 C_{ll^{\prime}}^{mn}=(-1)^n \frac{\Gamma(m+n+2l)}{\left(m+l^{\prime}\right)!l^{\prime}!\left(n+l-l^{\prime}\right)!\left(l-l^{\prime}\right)!}\;,
\end{equation}
with $C_{00}^{00}=0$.
\section{Self--energy in the case of fluctuating colloids}\label{app:se}
{For fluctuating colloids, the self--energy part of the multipole--multipole
interaction of the auxiliary fields $\Psi_{i}$ was introduced in Eq.~(\ref{Zkard3}).
Here we calculate it explicitly with a method similarly to the one employed in
Ref.~\cite{rods}. As the starting point, we obtain from Eqs.~(\ref{Zkard2}) and
(\ref{Zkard3}) the following expession for $\mathcal{Z}_{i,{\rm self}} =
\exp\left( -\frac{k_{\rm B}T}{2\gamma} \mc{H}_{i,\rm self}\right)$:}\\
\begin{eqnarray}
\mathcal{Z}_{i,{\rm self}} &=&
 \int \prod_{i=1}^2 \mc{D}\psi_i\,\delta
\left(\Psi_{im}^c -\int_{S_i}d^2x\,(\cos (m\xi) / \cosh (m\xi_0))
 \cos(m\eta)\psi({\bf x})\right) \nonumber\\
 && \times \,\delta
\left(\Psi_{im}^s -\int_{S_i}d^2x\,(\sinh (m\xi) / \sinh (m\xi_0))
 \sin(m\eta) \psi({\bf x})\right) 
\exp\left( {\rm i}\sum_m (\Psi_{im}^c P_{im} +\Psi_{im}^s Q_{im})  \right) \nonumber \\
&&\times \exp\left(
-\frac{k_{\rm B}T}{2\gamma}
\int_{S_i}d^2x \int_{S_i}d^2x'\,
\psi_i({\bf x})\,G(|{\bf x}-{\bf x'}|)\,\psi_i({\bf x})
-{\rm i}\int_{S_{i,\rm ref}}d^2x\,\psi_i({\bf x})\,f_{i,\rm ext}({\bf x})
\right)\;. \nonumber\\
\end{eqnarray}
The $\delta$--functions in $\mathcal{Z}_{i,{\rm self}}$ may be eliminated by 
introducing conjugate multipole moments $\widetilde{\Psi}_{im}^c$ and $\widetilde{\Psi}_{im}^s$
 to the multipoles $\Psi_{im}^{c \choose  s}$ of the auxiliary fields:
\begin{eqnarray}
\label{conjug}
&& \delta\left(\Psi_{im}^{c \choose s}-\int_{S_{i,{\mathrm {ref}}}}d^2x {\cosh (m\xi_i)\cos (m\eta_i)/\cosh (m\xi_0) \choose \sinh (m\xi_i)\sin (m \eta_i)/\sinh (m\xi_0)} \psi_i(\mathbf x_i)\right) = \nonumber \\
&& \int d\widetilde{\Psi}_{im}^{c \choose s} \exp\left( {\rm i}\widetilde{\Psi}_{im}^{c \choose s} \left[\Psi_{im}^{c \choose s}-\int_{S_{i,{\mathrm {ref}}}}d^2x\; {\cosh (m\xi_i)\cos (m\eta_i)/\cosh (m\xi_0) \choose \sinh (m\xi_i)\sin( m \eta_i)/\sinh (m\xi_0)} \psi_i(\mathbf x_i) \right]\right)\;.
\end{eqnarray}
Inserting Eq.~(\ref{conjug}) into $\mathcal{Z}_{i,{\rm self}}$ we obtain:
\begin{eqnarray}
\label{z-after-conjug}
{\mathcal Z}_{i, {\mathrm {self}}} &=& \int \prod_m d\widetilde{\Psi}_{im} \int \mathcal{D}\psi_i \exp\left(-\frac{k_{\mathrm B}T}{2\gamma} \int_{S_{i,{\mathrm{ref}}}} d^2x \int_{S_{i,\mathrm{ref}}} d^2x^{'} \psi_i(\mathbf x_i)G(|\mathbf x- \mathbf x^{'}|)\psi_i(\mathbf x_i)\right.\nonumber\\
&& \left. -{\rm i}\int_{S_{i,{\mathrm{ref}}}}d^2 x \;\psi_i(\mathbf x_i)\left[ \sum_m \frac{\cosh (m\xi_i)}{\cosh (m\xi_0)}(P_{im}+\widetilde{\psi}_{im}^c)\cos (m\eta_i) +\frac{\sinh (m\xi_i)}{\sinh (m\xi_0)}(Q_{im}+\widetilde{\psi}_{im}^s)\sin (m\eta_i)\right] \right. \nonumber\\
&&\left. +  {\rm i}\sum_{m=0}\left[(P_{im}+\widetilde{\Psi}_{im}^c)\Psi_{im}^c + (Q_{im}+\widetilde{\Psi}_{im}^s)\Psi_{im}^s\right]\right) \;,
\end{eqnarray}
where $d\widetilde{\Psi}_{im}=d\widetilde{\Psi}_{im}^s d\widetilde{\Psi}_{im}^c$. 
The functional integral $\int \mathcal{D} \psi_i$ in eq.~(\ref{z-after-conjug}) can be 
replaced by a functional integral over a constrained height field $h(\bf x)$:
\begin{eqnarray}
\label{replaced-z}
 {\mathcal Z}_{i, {\mathrm {self}}} &=& \int \prod_m d\widetilde{\Psi}_{im} \exp\left({\rm i}\sum_{m=0}^{\infty}\left[ (P_{im}+\widetilde{\Psi}_{im}^c)\Psi_{im}^c+(Q_{im}+\widetilde{\Psi}_{im}^s)\Psi_{im}^s\right]\right) \nonumber \\
&& \times \int \mathcal{D}h \prod_{\mathbf x_i \in S_{i, {\mathrm{ref}}}} \delta(h(\mathbf x_i)-\widetilde{f_i})\exp\left(-\frac{\gamma}{2k_{\mathrm B}T}\int d^2x\; \left[(\nabla h)^2 +\frac{h^2}{\lambda_c^2}\right]\right)\;,
\end{eqnarray}
where $\widetilde{f_i} = \sum_{m}\left[ \frac{\di \cosh (m\xi_i)}{\di \cosh (m\xi_0)}(P_{im}+\widetilde{\Psi}_{im}^c)\cos (m\eta_i) +\frac{\di \sinh (m\xi_i)}{\di \sinh (m\xi_0)}(Q_{im}+\widetilde{\Psi}_{im}^s)\sin (m\eta_i)\right] $.\\
{In the region $S_{i,{\rm ref}}$, i.e. the ellipse enclosed by the reference contact line, the height field
$h$ is pinned to $\widetilde{f_i}$. Therefore the contribution of the functional integral $\int \mathcal{D}h$ 
in this region is simply given by the surface energy of $\widetilde{f_i}$ which was determined in
Eq.~(\ref{energy_si}) and ${\mathcal Z}_{i, {\mathrm {self}}}$ becomes:} 
\begin{eqnarray}
\label{lambda->inf} 
{\mathcal Z}_{i, {\mathrm {self}}} &\stackrel{\lambda_c \rightarrow \infty}{\approx}& 
\int \prod_m d\widetilde{\Psi}_{im} \exp \left( -\frac{\gamma \pi}{2k_{\mathrm B}T} \sum_m m\left[
{(P_{im}+\widetilde{\Psi}_{im}^c)^2 \tanh (m\xi_0)} + {(Q_{im}+\widetilde{\Psi}_{im}^s)^2 \coth (m\xi_0)}\right] \right. 
\nonumber \\
&&\left. +{\rm i}\sum_m \left[ (P_{im}+\widetilde{\Psi}_{im}^c)\Psi_{im}^c+(Q_{im}+\widetilde{\Psi}_{im}^s)\Psi_{im}^s \right]\right) \nonumber\\
&& \times \int \mathcal{D}h \prod_{\mathbf x_i \in \partial S_{i, {\mathrm{ref}}}} \delta(h(\mathbf x_i)-\widetilde{f_i}) \exp \left(-\frac{\gamma}{2k_{\mathrm B}T}\int_{\mathbb{R}^2\setminus S_{i,{\rm ref}}} d^2x \; \left[(\nabla h)^2+\frac{h^2}{\lambda_c^2}\right] \right)\;,
\end{eqnarray}
where the remaining $\delta$-functions describe the pinning of $h(\bf x)$ to the boundaries 
$\partial S_{i, {\rm ref}}$ of the integration domain. The auxiliary field can be separated into two parts, 
$h = h_0 + h_1$, where $(-\nabla+\lambda_c^{-2})h_0=0$ with the boundary conditions 
$h_0({\bf x}_i)|_{\partial S_{i,{\rm ref}}}\equiv\widetilde{f_i}$ and 
$h_1({\bf x}_i)|_{\partial S_{i,{\rm ref}}}=0$. Applying Gauss' theorem to the integral in the 
exponent of Eq.~(\ref{lambda->inf}) leads to
\begin{eqnarray}
\label{z_h0+h1}
 {\mathcal Z}_{i, {\mathrm {self}}} &=& \int \prod_m d\widetilde{\Psi}_{im} \exp 
\left( -\frac{\gamma \pi}{2k_{\mathrm B}T} \sum_m m \left[{(P_{im}+\widetilde{\Psi}_{im}^c)^2 \tanh (m\xi_0)}+ 
{(Q_{im}+\widetilde{\Psi}_{im}^s)^2 \coth (m\xi_0)}\right] \right. \nonumber \\
&&\left. +{\rm i}\sum_m \left[ (P_{im}+\widetilde{\Psi}_{im}^c)\Psi_{im}^c+(Q_{im}+\widetilde{\Psi}_{im}^s)
\Psi_{im}^s \right]\right)
\exp\left(-\frac{\gamma}{2k_{\rm B}T}\oint d{\bf x}\; h_0({\bf x})\nabla h_0({\bf x})\right) \nonumber\\
&& \times \int \mathcal{D}h_1 \prod_{\mathbf x_i \in S_{i, {\mathrm{ref}}}} \delta(h_1(\mathbf x_i)) 
\exp \left(-\frac{\gamma}{2k_{\mathrm B}T}\int_{\mathbb{R}^2\setminus S_{i,{\rm ref}}} d^2x \; 
\left[(\nabla h_1)^2+\frac{h_1^2}{\lambda_c^2}\right] \right)\;.
\end{eqnarray}
The functional integral over $h_1$ yield to a constant value independent of any multipole moment, which can be neglected.\\
The general solution to the Helmholtz differential equation for $h_0$ in $\mathbb{R}^2 \setminus S_{i,{\rm ref}}$ 
is needed for computing the line integral in eq.~(\ref{z_h0+h1}). Separation of variables in the Helmholz 
equation in elliptic coordinates would lead to the Mathieu (angular part) and modified Mathieu (radial part) 
differential equations. The solution to these equations in the asymptotic case $\lambda_c \rightarrow \infty$ 
become standard triangular functions (angular part) and modified Bessel functions of the second kind
(radial part), respectively. Therefore, the solution is given by
\begin{equation}
\label{helm-sol}
h_0({\bf x}_i)=\sum_m \frac{K_m(a^{'}e^{\xi_i}/2\lambda_c)}{K_m(a^{'}e^{\xi_0}/2\lambda_c)}(A_m\cos (m\eta_i) + B_m\sin (m\eta_i)) \;.
\end{equation}
The coefficients in Eq.~(\ref{helm-sol}) are readily determined by comparing to the boundary conditions: 
$A_m=P_{im}+\widetilde{\Psi}_{im}^c$,  $B_m=Q_{im}+\widetilde{\Psi}_{im}^s$ and $B_0 = 0$. 
Then the line integral evaluates to $2\pi g(m) (A_m^2+B_m^2)$ with $g(m)=m/2\:(m>0)$ and 
$g(0) = -1/\ln(\gamma_{\rm e} a^{'} e^{\xi_0}/8\lambda_c)$. 
{Using this, ${\mathcal Z}_{i, {\mathrm {self}}}$ reads:}
\begin{eqnarray}
 {\mathcal Z}_{i, {\mathrm {self}}} &=& \int \prod_m d\widetilde{\Psi}_{im} \exp \left( 
-\frac{\gamma \pi}{2k_{\mathrm B}T} \sum_m m \left[{(P_{im}+\widetilde{\Psi}_{im}^c)^2 \tanh (m\xi_0)}+ 
{(Q_{im}+\widetilde{\Psi}_{im}^s)^2 \coth (m\xi_0)}\right] \right. \nonumber \\
&&\left. +{\rm i}\sum_m \left[(P_{im}+\widetilde{\Psi}_{im}^c)\Psi_{im}^c+
(Q_{im}+\widetilde{\Psi}_{im}^s)\Psi_{im}^s\right] \right) \\
&& \times \exp\left(-\frac{\gamma\pi}{k_{\rm B}T}\sum_{m=0} g(m)\left[(P_{im}+\widetilde{\Psi}_{im}^c)^2+(Q_{im}+\widetilde{\Psi}_{im}^s)^2\right]\right)\;. \nonumber
\end{eqnarray}
{The last integration over the conjugate multipole moments 
can be performed after shifting variables,
$\widetilde{\Psi}_{im}^c \to P_{im} + \widetilde{\Psi}_{im}^c$ and $\widetilde{\Psi}_{im}^s \to Q_{im} + 
\widetilde{\Psi}_{im}^c$. After this integration, the final result for   
$\mathcal{H}_{i,{\rm self}} =(-2\gamma)/(k_{\rm B}T) \ln {\mathcal Z}_{i, {\mathrm {self}}}$ is given by:}
\begin{eqnarray}
 \mathcal{H}_{i,{\rm self}} &=& - \frac{{\ln(\gamma_{\rm e} a^{'} e^{\xi_0}/8\lambda_c)}}{2\pi}{\Psi_{i0}^c}^2 \nonumber \\
&&+ \frac{1}{\pi}\sum_{m>0} \frac{1}{m}\left(\frac{{\Psi_{im}^c}^2}{1+\tanh (m\xi_0)}+\frac{{\Psi_{im}^s}^2}{1+\coth (m\xi_0)}\right)\;.
\end{eqnarray}
\end{small}

\end{document}